\documentclass[english,11pt]{article}
\usepackage{multicol}
\usepackage{amsmath}
\usepackage{amssymb}
\usepackage{bbold}
\usepackage{mathrsfs}
\usepackage{breqn}
\usepackage{graphicx}
\usepackage[table]{xcolor}
\usepackage{accents}
\usepackage{float}
\usepackage{color}
\usepackage[T1]{fontenc}
\usepackage[utf8]{inputenc}

\usepackage[colorlinks=true,urlcolor=blue,linkcolor=blue,citecolor=red]{hyperref}

\usepackage[
backend=bibtex,
style=numeric,
maxnames=3,
minnames=1,
sorting=none,
doi=false,
url=false
]{biblatex}

\renewbibmacro*{journal}{%
  \iffieldundef{shortjournal}
    {\printfield{journaltitle}}
    {\printfield{shortjournal}}}

\DeclareMathOperator{\co}{c} 
\DeclareMathOperator{\si}{s}
\DeclareMathOperator{\ch}{ch}
\DeclareMathOperator{\sh}{sh}

\begin{document}
\begin{center}
{\Large \bf Horizon Singularities in the Schwarzschild Geometry of the Teleparallel Equivalent of General Relativity\\ }
\vskip 0.7cm
{D. F. L\'opez\footnote{Corresponding author: diego.lopez@dal.ca} and A. A. Coley
}
\vskip 0.2cm
{\it Department of Mathematics and Statistics\\ 
Dalhousie University\\
Halifax, Canada}
\vskip 0.5cm
{R. J. van den Hoogen
}
\vskip 0.2cm
{\it Department of Mathematics and Statistics\\ 
St. Francis Xavier University\\
Antigonish, Canada}\\
\vskip 1cm
\begin{quote}
{\bf Abstract.}~{\small Certain torsion scalar invariants are known to diverge at the horizon of the Schwarzschild solution in the Teleparallel Equivalent of General Relativity (TEGR), obstructing its interpretation as a black hole spacetime. We show that Schwarzschild TEGR geometries split into two distinct subclasses determined by the Lorentz sector of the geometry---the elements of the tetrad and spin connection not encoded in the metric but appearing in the torsion. In the regular subclass, the divergences are absent and the horizon belongs to the manifold, supporting a consistent black hole interpretation. In the singular subclass, the divergences are genuine and the horizon is excluded from the manifold. As part of this analysis, we clarify the role of inertial contributions in teleparallel gravity, showing that the proper frame does not, by itself, eliminate inertial effects and that the horizon singularities are independent of the inertial structure of the frame. Using three independent approaches---the inertial frame condition, horizon-penetrating coordinates, and the horizon regularity criterion---we determine the complete class of Lorentz sector functions compatible with a regular horizon in Schwarzschild TEGR geometries. For this class, all torsion scalar invariants remain finite at the horizon and analytic extensions across it are admitted.}
\end{quote}
\end{center}
\vskip 1.0cm

\section{Introduction}

Teleparallel theories of gravity provide an alternative geometric formulation of gravitation, in which gravitational effects are encoded in the torsion of a flat connection rather than in the curvature of the Levi-Civita connection. Throughout this work, we use the term ``teleparallel'' to refer specifically to metric teleparallel theories and geometries. The geometric setup underlying these theories is known as teleparallel geometry (TG), which contrasts with the semi-Riemannian geometry (sRG) underlying curvature-based theories of gravity such as General Relativity (GR). Within the family of teleparallel theories of gravity, the Teleparallel Equivalent of General Relativity (TEGR) is distinguished by its local dynamical equivalence to GR, despite being formulated within TG rather than sRG. The Lagrangian density of TEGR is the torsion scalar $T$ of TG, which differs from the Ricci scalar $R$ of sRG only by a total divergence. This relation guarantees dynamical equivalence at the level of the field equations (FE) and the classical predictions of the theory~\cite{aldrovandi2012teleparallel,bahamonde2023teleparallel}.

Recent analyses of static, spherically symmetric vacuum teleparallel spacetimes have reaffirmed earlier results~\cite{obukhov2003metric} indicating that scalar invariants constructed from the torsion tensor can diverge at the location corresponding to the Schwarzschild horizon. In particular, for Schwarzschild solutions in TEGR, it has been shown that certain torsion scalars, including the torsion scalar $T$ itself, diverge at the putative horizon~\cite{coley2024spherically,van2024teleparallel,golovnev2024static}. Similar conclusions arise in the broader class of teleparallel theories known as $F(T)$ gravity, where the existence of a local horizon generically entails a divergence of the torsion scalar~\cite{coley2025black}. In New General Relativity (NGR), a teleparallel theory of gravity whose Lagrangian is the most general quadratic combination of torsion invariants and of which TEGR is a special case, physically viable models have likewise been found to exhibit divergences of torsion scalars at local horizons~\cite{lopez2025black}. This persistent behaviour has been suggested to reflect a structural feature of TG itself~\cite{coley2024spherically,golovnev2024static}, an interpretation we revisit in this work.

A divergence of a torsion scalar invariant signals a genuine singularity of the teleparallel geometric structure, preventing a smooth extension of the manifold through the horizon. Consequently, in such cases, the horizon and the region interior to it do not form part of the manifold. This raises a nontrivial interpretational challenge: if the metric solution coincides with the Schwarzschild solution and test particles follow the same metric geodesics as in GR, in what sense can the divergence of torsion invariants at the horizon be considered physically meaningful in TEGR?

Part of the difficulty stems from the additional geometric freedom present in TG: whereas sRG is fully specified by the ten independent components of the metric tensor, TG requires a tetrad with sixteen components, only ten of which are constrained by the metric. The remaining six components correspond to the local Lorentz orientation of the orthonormal frame at each point and constitute what we shall refer to as the Lorentz sector of the geometry, which we discuss in Section~\ref{LorentzSector}. In the static and spherically symmetric case, the most general teleparallel geometry is characterized by four independent functions~\cite{mcnutt2023frame}: two metric functions and two Lorentz sector functions~\cite{van2024teleparallel,mcnutt2023frame}. This introduces an important distinction: we must sometimes distinguish within an entire class of teleparallel geometries sharing a given metric and symmetry (e.g.\ all static and spherically symmetric geometries), while any particular representative $(\hat{h},\hat{\omega})$ within such a class is related to any other by a local Lorentz transformation.

In TEGR, the FE depend only on the metric and therefore determine only the metric functions, leaving the Lorentz sector entirely unconstrained by the dynamics. In more general teleparallel theories of gravity, such as NGR or $F(T)$ gravity, the FE generally depend explicitly on the Lorentz sector and may constrain it partially or completely~\cite{coley2025black,lopez2025black}. In this sense, TEGR constitutes a unique case among teleparallel theories, and it is the theory we focus on throughout this work. Although the dynamical predictions of TEGR coincide with those of GR, the teleparallel description introduces additional geometric structure that is imperceptible at the metric level, and this structure is central to the interpretation of the theory~\cite{weatherall2024general}. This situation raises the question of how the Lorentz sector of TEGR should be fixed, and what role it plays in the emergence and interpretation of the torsion singularities.

Addressing these questions requires two preliminary results. The first concerns the inertial structure of teleparallel frames. In sRG, the Levi-Civita connection encodes both gravitational and inertial effects in an inseparable manner. In contrast, in TG these contributions are commonly understood to split into the contortion tensor and the spin connection, respectively, with the spin connection interpreted as a pure gauge object representing inertial effects~\cite{krvsvsak2015spin,krvsvsak2019teleparallel}. In particular, one can transform to the so-called proper frame, in which the spin connection vanishes, which is often taken to imply that all inertial contributions have been removed. As we will show, this claim is not correct. In the static and spherically symmetric case, when one performs a Lorentz transformation to move to the proper frame, the freedom previously encoded in the spin connection is not eliminated but is simultaneously transferred to the transformed tetrad~\cite{van2024teleparallel,krvsvsak2024teleparallel}. As a result, the equation of motion in the proper frame still contains the contortion tensor and does not reduce to the special-relativistic free-fall form: the inertial contributions have not been eliminated in any fundamental sense. We provide a precise characterization of inertial contributions in teleparallel theories, based on the well-established sRG notion of Fermi-Walker transport~\cite{wald2010general,misner1973gravitation}, define the corresponding inertial frame condition, and show that the proper frame does not in general satisfy it.

The second preliminary result concerns the relationship between inertial effects and the Lorentz sector. Since the torsion scalar is invariant under local Lorentz transformations, it is unaffected by the Lorentz transformations that implement the inertial frame condition. The Lorentz sector of the geometry and the inertial structure of the frame are therefore two independent and distinct structures within the teleparallel description. In particular, the dependence of the torsion scalar on the Lorentz sector does not reflect inertial contamination in the dynamical sense~\cite{krvsvsak2015spin,krvsvsak2019teleparallel}, and the horizon singularities cannot be attributed to inertial effects. They are instead, as we show, features of a specific subclass of inequivalent teleparallel geometries within the TEGR Schwarzschild family, while the complementary subclass is compatible with a regular horizon description.

In this work, we revisit static, spherically symmetric vacuum solutions in TEGR, focusing on torsion scalar invariants and their behaviour at the Schwarzschild horizon. We pursue three complementary approaches to resolving the horizon singularities: (i) the inertial frame condition, which provides a physically motivated selection of the Lorentz sector, (ii) horizon-penetrating coordinates, through which horizon-penetrating teleparallel geometries are constructed via coordinate and local Lorentz transformations, and (iii) the horizon regularity criterion, which provides a direct mathematical characterization of the general class of Lorentz sector function values compatible with a regular horizon. Our aim is to demonstrate that the Schwarzschild solution in TEGR admits a consistent black hole interpretation once the Lorentz sector is chosen to be compatible with a regular horizon description.

\section{Teleparallel geometry}

The fundamental geometric objects of TG are the orthonormal tetrad field $h^{a}{}_{\mu}$ and the metric-compatible (i.e., $\omega_{(ab)\mu}=0$), flat spin connection $\omega^{a}{}_{b\mu}$. The tetrad relates the spacetime metric $g_{\mu\nu}$ to the Minkowski metric in the tangent space $\eta_{ab} = \mathrm{diag}(-1,1,1,1)$ through
\begin{equation}\label{htog}
g_{\mu\nu} = h^{a}{}_{\mu}h^{b}{}_{\nu} \eta_{ab},
\end{equation}
while the spin connection is defined in terms of local Lorentz transformations $\Lambda^{a}{}_{b}=\Lambda^{a}{}_{b}(x)$ as
\begin{equation}\label{spin}
\omega^{a}{}_{b\mu}=\Lambda^{a}{}_{c}\partial_{\mu}\Lambda_{b}{}^{c},
\end{equation}
where $\Lambda^{a}{}_{c}\Lambda_{b}{}^{c}=\delta^{a}_{b}$ is the Lorentz orthogonality condition~\cite{krvsvsak2019teleparallel}. Together, the pair $(h^{a}{}_{\mu}, \omega^{a}{}_{b\mu})$ fully determines the teleparallel geometry.

The spin connection appears in the teleparallel covariant derivative, $\nabla_{\mu}$, acting on tangent-space indices and is related to the teleparallel connection $\Omega^{\alpha}{}_{\mu\nu}$, which acts only on spacetime indices~\cite{aldrovandi2012teleparallel}, through
\begin{equation}
\omega^{a}{}_{b\mu}\equiv h^{a}{}_{\nu}\nabla_{\mu}h_{b}{}^{\nu}
= h^{a}{}_{\nu}\partial_{\mu}h_{b}{}^{\nu}
+ h^{a}{}_{\alpha}\Omega^{\alpha}{}_{\nu\mu}h_{b}{}^{\nu}.
\end{equation}
Inverting this relation yields the teleparallel connection in terms of the tetrad and spin connection,
\begin{equation}\label{Omega}
\Omega^{\rho}{}_{\nu\mu}
= h_{a}{}^{\rho} \partial_{\mu} h^{a}{}_{\nu}
+ h_{a}{}^{\rho} \omega^{a}{}_{b\mu} h^{b}{}_{\nu}\,.
\end{equation}
Since this connection is flat by construction, its curvature vanishes identically, and the only non-trivial field strength is the torsion tensor, given by the antisymmetric part of the teleparallel connection,
\begin{equation}\label{torsion}
T^{\sigma}{}_{\mu\nu} = 2\Omega^{\sigma}{}_{[\nu\mu]}\,.
\end{equation}

The torsion tensor~\eqref{torsion} admits a decomposition into three Lorentz-irreducible components, conventionally referred to as the \textit{vector}, \textit{axial}, and purely \textit{tensorial} parts~\cite{bahamonde2023teleparallel,
hayashi1979new}:
\begin{equation}\label{TorDec}
\mathscr{V}_{\mu} = T^{\nu}{}_{\nu\mu}\,,\quad
\mathscr{A}_{\mu} = \frac{1}{6}\varepsilon_{\mu\nu\rho\sigma}T^{\nu\rho\sigma}\,,\quad
\mathscr{T}_{\sigma\mu\nu} = T_{(\sigma\mu)\nu}
+ \frac{1}{3}\left(g_{\sigma[\nu}\mathscr{V}_{\mu]}
+ g_{\mu[\nu}\mathscr{V}_{\sigma]}\right),
\end{equation}
where $\varepsilon_{\mu\nu\rho\sigma}$ denotes the totally antisymmetric Levi-Civita tensor associated with the metric $g_{\mu\nu}$. Each of these components defines a scalar invariant through full contraction,
\begin{equation}\label{AVTS}
\mathscr{A} = \mathscr{A}^{\mu}\mathscr{A}_{\mu},
\qquad
\mathscr{V} = \mathscr{V}^{\mu}\mathscr{V}_{\mu},
\qquad
\mathscr{T} = \mathscr{T}^{\sigma\mu\nu}\mathscr{T}_{\sigma\mu\nu}.
\end{equation}
These scalar invariants appear in the gravitational Lagrangian densities of a variety of teleparallel theories of gravity, including TEGR, NGR, and $F(T)$ gravity~\cite{bahamonde2023teleparallel,hayashi1979new}. In particular, they combine to form the torsion scalar~\cite{aldrovandi2012teleparallel},
\begin{equation}\label{TS}
T = \frac{3}{2}\,\mathscr{A}
- \frac{2}{3}\,\mathscr{V}
+ \frac{2}{3}\,\mathscr{T},
\end{equation}
which constitutes the gravitational part of the TEGR Lagrangian density and is equivalent to the Ricci scalar \(R\) of sRG up to a boundary term~\cite{aldrovandi2012teleparallel}.

\subsection{Connection decomposition}

A fundamental result of teleparallel geometry, as a consequence of the Ricci theorem~\cite{krvsvsak2019teleparallel,kobayashi1996foundations}, is that the teleparallel spin connection $\omega^{a}{}_{b\mu}$ can be decomposed into two geometrically distinct contributions. The first is the Levi-Civita spin connection\footnote{Also known as the Ricci rotation coefficients.}
\begin{equation}
\gamma^{a}{}_{b\mu}\equiv h^{a}{}_{\nu}D_{\mu}h_{b}{}^{\nu}
= h^{a}{}_{\nu}\partial_{\mu}h_{b}{}^{\nu}
+ h^{a}{}_{\alpha}\Gamma^{\alpha}{}_{\nu\mu}h_{b}{}^{\nu},
\end{equation}
where $D_{\mu}$ is the covariant derivative with respect to the Levi-Civita connection given by
\begin{equation}\label{Gamma}
\Gamma^{\alpha}{}_{\nu\mu}
=\frac{1}{2}g^{\alpha\lambda}
\left(\partial_{\mu}g_{\lambda\nu}
+\partial_{\nu}g_{\lambda\mu}
-\partial_{\lambda}g_{\nu\mu}\right).
\end{equation}
The second contribution is the contortion tensor
\begin{equation}
K_{ab\mu}=\frac{1}{2}T_{\mu ab}-T_{[ab]\mu},
\end{equation}
which is antisymmetric in its first two indices, $K_{(ab)\mu}=0$. This decomposition can be written as~\cite{aldrovandi2012teleparallel}
\begin{equation}\label{conth}
\gamma^{a}{}_{b\mu}=\omega^{a}{}_{b\mu}-K^{a}{}_{b\mu},
\end{equation}
or equivalently
\begin{equation}\label{conth2}
\Gamma^{\rho}{}_{\mu\nu}=\Omega^{\rho}{}_{\mu\nu}-K^{\rho}{}_{\mu\nu}.
\end{equation}
This identity allows the Ricci scalar of sRG to be expressed in terms of the torsion scalar of TG up to a total divergence. A direct calculation using \eqref{conth} yields~\cite{aldrovandi2012teleparallel,bahamonde2023teleparallel}
\begin{equation}\label{equiv}
R=-T-\frac{2}{h}\partial_{\nu}\left(h\,\mathscr{V}^{\nu}\right),
\end{equation}
where $R$ is the Ricci scalar of the Levi-Civita connection and $h$ is the tetrad determinant. Since the divergence term does not contribute to the FE, Eq.~\eqref{equiv} establishes the dynamical equivalence between GR and TEGR.

\subsection{Spacetime notion}\label{LorentzSector}

In an sRG description of spacetime, the geometry is specified by a metric tensor ($g_{\mu\nu}$), with ten independent components, defined on a four-dimensional manifold ($\mathcal{M}$)~\cite{wald2010general}. In this framework, spacetime is characterized by
\begin{equation}
\text{sRG spacetime:}\;\; (\mathcal{M}, g).
\end{equation}
Spacetime consists of all regular events, excluding genuine singularities, which correspond to a breakdown of the geometrical structure and therefore lie outside $\mathcal{M}$. While singularities are fundamentally characterized by geodesic incompleteness, the divergence of curvature scalar invariants provides a sufficient indication of their presence~\cite{wald2010general,hawking2023large}.

In contrast, the TG structure is determined by both an orthonormal tetrad ($h^{a}{}_{\mu}$) and a flat, metric-compatible spin connection ($\omega^{a}{}_{b\mu}$)~\cite{maluf2013teleparallel,krvsvsak2016covariant}. The tetrad field possesses sixteen independent components~\cite{krvsvsak2019teleparallel}, of which only ten are constrained by the metric through~\eqref{htog}. The remaining six represent the local Lorentz orientation of the orthonormal frame. In the covariant formulation, this information can be distributed between the tetrad and the spin connection, and constitute the \emph{Lorentz sector} of the teleparallel geometry. The Lorentz sector is not visible at the metric level: different values of the Lorentz sector functions yield the same metric. However, the Lorentz sector functions appear in the torsion tensor, $T^{\alpha}{}_{\mu\nu}$, and in all torsion-derived quantities, including all torsion scalar invariants, and therefore distinguish inequivalent teleparallel geometries associated with the same metric structure. The teleparallel geometry therefore contains additional structure beyond the metric description~\cite{weatherall2024general,krvsvsak2024teleparallel}, and spacetime is characterized by
\begin{equation}
\text{TG spacetime:}\;\; (\mathcal{M}, h, \omega).
\end{equation}
Spacetime consists of all regular events, while genuine geometric singularities correspond to a breakdown of the teleparallel structure and therefore lie outside the manifold. A sufficient indication of such a breakdown is the divergence of torsion scalar invariants, although this need not provide an exhaustive characterization of teleparallel singularities.

This richer geometric structure has direct consequences for the physical interpretation of torsion scalar invariants. In sRG, curvature scalar invariants such as the Ricci scalar are uniquely determined by the metric and therefore carry no information beyond it~\cite{wald2010general}. In the teleparallel description, by contrast, torsion scalar invariants depend not only on the metric but also on the Lorentz sector. Consequently, distinct values of the Lorentz sector produce different torsion scalar invariants while leaving all metric observables unchanged. This is a genuine feature of the teleparallel framework: torsion scalar invariants encode geometric information beyond that contained in the metric alone. As a result, different Lorentz sectors correspond to qualitatively distinct teleparallel geometries.

\subsection{Equation of motion}

In sRG, geodesics are curves whose tangent vector $u^{\mu}$ is parallel transported along itself with respect to the Levi--Civita connection,
\begin{equation}\label{Geo2}
u^{\mu}D_{\mu}u^{\rho}\equiv\frac{d u^{\rho}}{d\tau} + \Gamma^{\rho}{}_{\mu\nu}\, u^{\mu} u^{\nu} = 0.
\end{equation}
In sRG-based theories where matter couples minimally to the spacetime metric, the equation of motion~\eqref{Geo2} follows from extremizing the action
\begin{equation}\label{testp}
S=-m\int \sqrt{g_{\mu\nu}\,dx^{\mu}dx^{\nu}},
\end{equation}
so that the resulting trajectories are entirely determined by the metric structure of spacetime, independently of any additional geometric variables.

In TG-based theories with minimal coupling, the corresponding equation of motion coincides with~\eqref{Geo2}. Using the decomposition~\eqref{conth2}, it can be rewritten in teleparallel form as~\cite{aldrovandi2012teleparallel,aldrovandi2008inertia}
\begin{equation}\label{Eqm2}
\frac{d u^{\rho}}{d\tau} + \left(\Omega^{\rho}{}_{\mu\nu}-K^{\rho}{}_{\mu\nu}\right)\, u^{\mu} u^{\nu} = 0.
\end{equation}
Since the physical trajectories are still entirely determined by the metric, the Lorentz sector does not affect the motion, and Eqs.~\eqref{Geo2} and~\eqref{Eqm2} describe the same class of curves. Although Eq.~\eqref{Eqm2} can be written in a force-like form in the teleparallel context, with the contortion tensor playing the role of a gravitational force~\cite{aldrovandi2012teleparallel,maluf2013teleparallel}, it remains mathematically equivalent to the geodesic equation~\eqref{Geo2}.

In an orthonormal frame, the geodesic equation~\eqref{Geo2} can be expressed in terms of frame indices as
\begin{equation}\label{Geo}
\frac{d u^{a}}{d\tau} + \gamma^{a}{}_{b\mu}\, u^{b} u^{\mu} = 0,
\end{equation}
where $u^{a}=h^{a}{}_{\mu}u^{\mu}$. Similarly, in the teleparallel framework, the equation of motion~\eqref{Eqm2} takes the form
\begin{equation}\label{Eqm}
\frac{d u^{a}}{d\tau} + \bigl(\omega^{a}{}_{b\mu} - K^{a}{}_{b\mu}\bigr)\, u^{b} u^{\mu} = 0.
\end{equation}
Alternatively, Eq.~\eqref{Eqm} can be obtained directly from Eq.~\eqref{Geo} by using the decomposition~\eqref{conth}. The frame formulation is particularly useful because it isolates the physically relevant components of the connection associated with a given observer, in contrast to the coordinate expression, which depends on the choice of chart.

\subsection{Local inertial coordinates}

In sRG, at any spacetime point $p\in\mathcal{M}$, one can always introduce a coordinate system $x^{\mu}$ in which the metric takes its Minkowski form,
\begin{equation}\label{gpoint}
g_{\mu\nu}(p)=\eta_{\mu\nu},
\end{equation}
and its first derivatives vanish. This is equivalent to the vanishing of the Levi--Civita connection at $p$,
\begin{equation}\label{Gpoint}
\Gamma^{\rho}{}_{\mu\nu}(p)=0,
\end{equation}
although, in general, the second derivatives of the metric cannot be made to vanish simultaneously~\cite{wald2010general,misner1973gravitation}. Such coordinates are known as \emph{local inertial coordinates}, and they realize the standard construction of Riemann normal coordinates centred at $p$~\cite{misner1973gravitation}. In these coordinates, the spacetime metric coincides with that of flat Minkowski space up to first order in a neighbourhood of $p$, providing a precise formulation of the statement that sufficiently small regions of spacetime appear locally flat.

At $p$, the geodesic equation~\eqref{Geo2} reduces to
\begin{equation}\label{eqzero2}
\frac{d u^{\rho}}{d\tau} = 0,
\end{equation}
so that the motion locally coincides with inertial motion in special relativity (SR). This construction is intrinsically local: although Eqs.~\eqref{gpoint} and~\eqref{Gpoint} can always be achieved at a single event, they cannot in general be extended to a finite neighbourhood of $p$, since spacetime curvature obstructs the simultaneous vanishing of the second derivatives of the metric.

In TG, it is likewise always possible to introduce, at each point, an orthonormal basis in which the metric satisfies~\eqref{gpoint}. In particular, if Eq.~\eqref{gpoint} holds at $p$ and the relation~\eqref{htog} is satisfied, the tetrad can be chosen such that
\begin{equation}
h^{a}{}_{\mu}(p)=\delta^{a}_{\mu},
\end{equation}
which is equivalent to Eq.~\eqref{gpoint}. Using the decomposition~\eqref{conth2}, the condition~\eqref{Gpoint} becomes
\begin{equation}
\Omega^{\rho}{}_{\mu\nu}(p)-K^{\rho}{}_{\mu\nu}(p)=0,
\end{equation}
which is the natural teleparallel analog of~\eqref{Gpoint}. This shows that the contributions from the teleparallel connection and the contortion tensor balance exactly, so that even in local inertial coordinates the inertial and gravitational information encoded in these quantities does not vanish individually, but only through their combination. As a consequence, the equation of motion~\eqref{Eqm2} reduces to~\eqref{eqzero2}, recovering inertial motion locally in agreement with SR.

\subsection{Inertial comoving frames}\label{inerob}

It is essential to distinguish clearly between the effects of choosing coordinates and those arising from the choice of frames. Coordinates are merely labels used to parametrize spacetime points and carry no direct physical meaning. In contrast, a frame is a genuine geometric object corresponding to a choice of local basis in the tangent space, and it is this basis that encodes how physical measurements are defined and performed~\cite{misner1973gravitation,maluf2013teleparallel,maluf2007reference}.

A \emph{comoving frame} is the local orthonormal tetrad carried by an observer whose timelike basis vector (labelled by the index $1$) coincides with the observer's four-velocity,
\begin{equation}\label{comoving}
u^{\mu}\equiv h_{1}{}^{\mu}.
\end{equation}
Such a frame exists for any observer, regardless of whether the motion is geodesic or accelerated, and its construction is purely kinematical. Although the tetrad is orthonormal at each event, it does not in general eliminate the observer's proper acceleration or the rotation of the spatial triad~\cite{misner1973gravitation}.

An \emph{inertial comoving frame} is a comoving frame carried by a freely falling observer and transported without intrinsic rotation along the worldline. To make this precise, define the projected Ricci rotation coefficients
\begin{equation}
\Sigma_{ab}\equiv\gamma_{ab\mu}\,u^{\mu},
\end{equation}
which measure the rate of change of the tetrad along the observer's trajectory. Their boost components encode the proper acceleration, $a^{a}=\Sigma^{a}{}_{b}u^{b}$, while their spatial components describe the angular velocity of the spatial triad, $\sigma^{a}=\frac{1}{2}\epsilon^{a}{}_{bcd}u^{b}\Sigma^{cd}$~\cite{misner1973gravitation,maluf2013teleparallel,maluf2007reference}. In sRG, the inertial  frame condition, is imposed by requiring the tetrad to undergo Fermi--Walker transport along the observer's four-velocity. For geodesic motion, this reduces to parallel transport, yielding
\begin{equation}\label{sRGinertial}
\Sigma_{ab}=0.
\end{equation}
This does not require the full spin connection $\gamma^{a}{}_{b\mu}$ to vanish globally; only its contraction with $u^{\mu}$ must vanish~\cite{wald2010general,misner1973gravitation}. When condition~\eqref{sRGinertial} is satisfied, the geodesic equation~\eqref{Geo} reduces to
\begin{equation}\label{eqzero}
\frac{d u^{a}}{d\tau}=0,
\end{equation}
recovering the local special-relativistic form of inertial motion. In contrast to the pointwise construction of local inertial coordinates discussed in the preceding subsection, this provides a \emph{worldlinewise} realization of inertial motion: the connection coefficients vanish along the entire trajectory of the freely falling observer, rather than only at a single event~\cite{misner1973gravitation}.

Using identity~\eqref{conth}, the same physical requirement translates into TG as
\begin{equation}\label{inertial}
\Sigma_{ab}\equiv\bigl(\omega_{ab\mu}-K_{ab\mu}\bigr)u^{\mu}=0.
\end{equation}
This condition states that, along the observer's direction, the inertial contributions encoded in the spin connection exactly cancel those arising from the contortion tensor, according to the standard decomposition~\cite{krvsvsak2019teleparallel}. Any tetrad--spin connection pair satisfying~\eqref{comoving} and~\eqref{inertial} therefore defines an inertial comoving frame in TG. This definition is coordinate invariant, but it is not invariant under local Lorentz transformations. Under such a transformation, $\Sigma^{a}{}_{b}$ transforms as
\begin{equation}\label{Sigmatrans}
{\Sigma'}^{a}{}_{b}=\Lambda_{1}{}^{1}\Lambda^{a}{}_{c}\Lambda_{b}{}^{d}\Sigma^{c}{}_{d}+\Lambda^{a}{}_{c}\left(\Lambda_{1}{}^{e}h_{e}{}^{\mu}\partial_{\mu}\Lambda_{b}{}^{c}+\gamma^{c}{}_{d\mu}\Lambda_{b}{}^{d}\Lambda_{1}{}^{i}h_{i}{}^{\mu}\right),
\end{equation}
where $i\in\{2,3,4\}$ labels the spatial components of the original frame. The condition $\Sigma^{a}{}_{b}=0$ is therefore preserved only under Lorentz transformations satisfying
\begin{equation}\label{conG}
{u'}^{\mu}\partial_{\mu}\Lambda_{b}{}^{c}+\Lambda_{b}{}^{d}\gamma^{c}{}_{d\mu}v^{\mu}=0,
\end{equation}
where ${u'}^{\mu}=\Lambda_{1}{}^{e}h_{e}{}^{\mu}$ and $v^{\mu}=\Lambda_{1}{}^{i}h_{i}{}^{\mu}$. The first term in~\eqref{conG} is the directional derivative of $\Lambda_{b}{}^{c}$ along the transformed four-velocity ${u'}^{\mu}$, while the second term reflects the contribution from the spatial directions $v^{\mu}$ arising from the mixing introduced by the Lorentz transformation. The transformed four-velocity decomposes accordingly as
\[
{u'}^{\mu}=\Lambda_{1}{}^{1}u^{\mu}+v^{\mu},
\]
separating the boost along the original worldline from the spatial frame mixing. The vanishing of the sum in~\eqref{conG} therefore characterizes those local Lorentz transformations that map one inertial comoving frame into another, thereby identifying the residual freedom of the inertial comoving frame condition. When the transformation does not mix timelike and spatial directions, $\Lambda_{1}{}^{i}=0$, the transformed four-velocity coincides with the original one, and the transformation rule~\eqref{Sigmatrans} reduces to
\begin{equation}
{\Sigma'}^{a}{}_{b}=\Lambda^{a}{}_{c}\Lambda_{b}{}^{d}\Sigma^{c}{}_{d}+\Lambda^{a}{}_{c}\,u^{\mu}\partial_{\mu}\Lambda_{b}{}^{c},
\end{equation}
while condition~\eqref{conG} becomes
\begin{equation}
u^{\mu}\partial_{\mu}\Lambda_{b}{}^{c}=0.
\end{equation}
Thus, the Lorentz transformation must remain constant along the observer's worldline. More generally, when the Lorentz transformation changes the observer congruence, the variation of $\Lambda$ along the transformed worldline must compensate the contribution inherited from the original frame projected onto the new worldline, precisely as expressed by~\eqref{conG}.

\subsection{Revisiting the proper frame}\label{rproper}

In TG, a proper frame is often described as a frame in which inertial effects are absent, characterized by the vanishing of the teleparallel spin connection~\cite{krvsvsak2015spin,krvsvsak2019teleparallel,aldrovandi2008inertia,lucas2009regularizing}. We now examine whether this interpretation is physically justified.

In sRG, the covariant derivative of the frame $h_{a}{}^{\nu}$ along the observer's direction $u^{\mu}$ is given by~\cite{misner1973gravitation}
\begin{equation}
u^{\mu} D_{\mu} h_{b}{}^{\nu}=\Sigma^{a}{}_{b}\, h_{a}{}^{\nu}.
\end{equation}
When condition~\eqref{sRGinertial} is satisfied, the geodesic equation~\eqref{Geo} reduces to~\eqref{eqzero}, thereby locally recovering the special-relativistic form of inertial motion.

In TG, one may attempt to construct an analogous procedure by noting that the covariant derivative of the frame $h_{b}{}^{\nu}$ along $u^{\mu}$ with respect to the teleparallel connection is given by
\begin{equation}
u^{\mu} \nabla_{\mu} h_{b}{}^{\nu}
=\Xi^{a}{}_{b}\, h_{a}{}^{\nu}, \quad \Xi^{a}{}_{b}\equiv \omega^{a}{}_{b\mu} u^{\mu},
\end{equation}
and one may propose the condition
\begin{equation}\label{inerfake} 
\Xi^{a}{}_{b}=0,
\end{equation}
as the criterion for a frame free of inertial effects. This condition states that the frame is parallel transported with respect to the teleparallel connection along the observer's direction, formally mirroring the sRG construction. However, this requirement is not aligned with the equation of motion, since
\begin{equation}
\Xi^{a}{}_{b}u^{b}=a^{a}+K^{a}{}_{b\mu}u^{b}u^{\mu}\neq a^{a}.
\end{equation}
The condition $\Xi^{a}{}_{b}=0$ therefore requires $K^{a}{}_{b\mu}\, u^{b} u^{\mu}=0$, which does not hold in general. Moreover, no component of $\Xi^{a}{}_{b}$ alone encodes the proper acceleration $a^{a}$. Consequently, when a frame satisfies~\eqref{inerfake}, the equation of motion~\eqref{Eqm} reduces to
\begin{equation}
\frac{d u^{a}}{d\tau} = K^{a}{}_{b\mu}\, u^{b} u^{\mu},
\end{equation}
rather than recovering~\eqref{eqzero}. The observer is therefore not in free fall, and the resulting motion does not correspond to the inertial motion of SR.

This shows that~\eqref{inerfake} is not the appropriate criterion for defining inertial frames in TG. The inertial frame condition must be formulated consistently with the equation of motion: only in this way can one remove physical inertial effects, rather than merely describing transport relative to a chosen connection. Hence, requiring $\omega^{a}{}_{b\mu}u^{\mu}=0$ fails to characterize a frame in which inertial effects are absent, and consequently the stronger condition $\omega^{a}{}_{b\mu}=0$ also fails to do so. The claim that the proper frame is free of inertial effects is therefore incorrect.

A second and complementary perspective is provided by examining how the equation of motion~\eqref{Eqm} transforms when one moves to the proper frame. Let $\Lambda^{a}{}_{b}$ be a local Lorentz transformation such that
\begin{equation}\label{trans}
{h'}^{a}{}_{\mu}=\Lambda^{a}{}_{b}h^{b}{}_{\mu}, \qquad {\omega'}^{a}{}_{b\mu}= \Lambda^{a}{}_{c}\,\omega^{c}{}_{d\mu}\,\Lambda_{b}{}^{d}+\Lambda^{a}{}_{c}\,\partial_{\mu}\Lambda_{b}{}^{c}=0.
\end{equation}
Writing the equation of motion~\eqref{Eqm} in the proper frame yields
\begin{equation}
\frac{d {u'}^{a}}{d\tau} -{K'}^{a}{}_{b\mu}\, {u'}^{b} u^{\mu} =\Lambda^{a}{}_{b}\left(\frac{d {u}^{b}}{d\tau} + (\omega^{b}{}_{c\mu}-K^{b}{}_{c\mu})u^{c}u^{\mu}\right)=0,
\end{equation}
which simply reflects the covariant behaviour of the equation of motion under Lorentz transformations (${a'}^{a}=\Lambda^{a}{}_{b}a^{b}$). Passing to the proper frame therefore achieves only a redistribution of inertial contributions: they are removed from the spin connection and encoded in the transformed tetrad~\cite{van2024teleparallel,krvsvsak2024teleparallel}. The equation of motion in the proper frame still contains the contortion tensor and does not reduce to the special-relativistic free-fall form, so the proper frame possesses no special dynamical features.

The proper frame is therefore not the appropriate tool for removing inertial effects. Instead, one should note that in TG the equation of motion is governed by both the teleparallel spin connection and the contortion tensor, and the definition of inertial frames must be constructed consistently with that structure. The structural correspondence between sRG and TG is made explicit by displaying the geodesic equation and the inertial frame condition side by side
\begin{equation}
\text{sRG:}\quad u^{\nu}D_{\nu}u^{a}=0, \quad \Sigma_{ab}=0,
\end{equation}
\begin{equation}\label{condk}
\text{TG:}\quad u^{\nu}\nabla_{\nu}u^{a}=K^{a}{}_{b\nu}u^{b}u^{\nu}, \quad \Xi^{a}{}_{b}= K^{a}{}_{b\mu}u^{\mu}.
\end{equation}
These are related by identity~\eqref{conth}, since
\begin{equation}
\Sigma^{a}{}_{b}=\Xi^{a}{}_{b}-K^{a}{}_{b\mu}u^{\mu}.
\end{equation}
If one chooses to work in the proper frame while eliminating inertial effects, condition~\eqref{inertial} (or equivalently~\eqref{condk}) must still be imposed. This amounts to fixing the parameters of the Lorentz transformation used to reach the proper frame such that $\Sigma_{ab}=0$, whenever such freedom is available.

We therefore introduce the following definition: an \emph{inertial proper frame} is a comoving frame, $h_{1}{}^{\mu}=u^{\mu}$, in which simultaneously $\omega^{a}{}_{b\mu}=0$ and $\Sigma_{ab}=0$. The first condition sets the frame in the proper form; the second ensures that inertial effects are genuinely absent along the observer's worldline. Achieving an inertial proper frame is not always possible: it generally requires constraining the Lorentz sector functions, and when such freedom is unavailable the inertial proper frame is in general unattainable. For this reason the covariant approach is the more natural framework~\cite{krvsvsak2019teleparallel}.

In an inertial comoving frame, local inertial contributions vanish along the observer's worldline because the tetrad transport is consistent with the equation of motion. Gravitational effects persist through spacetime curvature in sRG and through torsion in TG, but they do not manifest as local forces on a freely falling observer. In both setups, when the inertial comoving condition~\eqref{sRGinertial} (or equivalently~\eqref{inertial}) is satisfied, the equation of motion reduces to~\eqref{eqzero}, thereby recovering the local special-relativistic form of inertial motion.

\section{Static spherically symmetric teleparallel geometries}

We work in standard spherical coordinates $x^{\mu}=(t,r,\theta,\phi)$ and assume spherical symmetry acting on two-dimensional spatial surfaces. The symmetry is generated by three rotational affine symmetry vector fields $X_{1}$, $X_{2}$, and $X_{3}$~\cite{coley2024spherically,mcnutt2023frame}, given explicitly by
\begin{subequations}\label{SS_generators}
\begin{align}
X_{1} &= \sin\phi\,\partial_{\theta} + \cos\phi\,\cot\theta\,\partial_{\phi}, \\
X_{2} &= -\cos\phi\,\partial_{\theta} + \sin\phi\,\cot\theta\,\partial_{\phi}, \\
X_{3} &= \partial_{\phi}.
\end{align}
\end{subequations}
The additional affine symmetry vector field $X_{4} = \partial_{t}$ encodes the assumption that the geometry is static. Under these symmetry requirements, the most general static, spherically symmetric teleparallel geometry~\cite{mcnutt2023frame} can be described by the diagonal orthonormal coframe
\begin{equation}\label{TetradSS}
h^{a}{}_{\mu} =
\begin{pmatrix}
A_{1} & 0 & 0 & 0 \\
0 & A_{2} & 0 & 0 \\
0 & 0 & r & 0 \\
0 & 0 & 0 & r\sin\theta
\end{pmatrix},
\end{equation}
together with the associated spin connection whose nonzero components are
\begin{equation}\label{SpinV}
\begin{gathered}
\omega_{41\theta} = \omega_{13\phi}/\sin\theta = \sin\chi \sinh\psi,\quad
\omega_{13\theta} = \omega_{14\phi}/\sin\theta = \cos\chi \sinh\psi, \\[1ex]
\omega_{42\theta} = \omega_{23\phi}/\sin\theta = \sin\chi \cosh\psi, \quad
\omega_{23\theta} = \omega_{24\phi}/\sin\theta = \cos\chi \cosh\psi, \\[1ex]
\omega_{21r} = \psi',
\quad
\omega_{43r} = \chi', \quad
\omega_{43\phi} = \cos\theta,
\end{gathered}
\end{equation}
where the coordinate freedom in $r$ has been used to fix the area-radius gauge, in which the radial coordinate satisfies $h^{3}{}_{\theta}=r$ and $h^{4}{}_{\phi}=r\sin\theta$. The metric functions are $A_{1}=A_{1}(r)$ and $A_{2}=A_{2}(r)$, and the Lorentz sector is parametrized by $\psi=\psi(r)$ and $\chi=\chi(r)$. The coframe~\eqref{TetradSS} and the spin connection~\eqref{SpinV} were first obtained in~\cite{mcnutt2023frame} and subsequently employed in~\cite{coley2024spherically,van2024teleparallel}.

\subsection{Inertial comoving frames}

To assign a physical interpretation to the tetrad in~\eqref{TetradSS}, we associate it with a family of observers by identifying the timelike component with the observer four-velocity as in~\eqref{comoving}. For the coframe given in~\eqref{TetradSS}, the associated observer four-velocity corresponds to a family of static observers,
\begin{equation}\label{fourv}
u^{\mu}=h_{1}{}^{\mu}=\left(\frac{1}{A_{1}},\,0,\,0,\,0\right).
\end{equation}
We now determine under which conditions the coframe~\eqref{TetradSS} represents an inertial comoving frame by evaluating condition~\eqref{inertial}. The spin connection contribution vanishes identically, $\omega_{ab\mu}u^{\mu}=0$, as follows directly from~\eqref{SpinV} and~\eqref{fourv}. Since this contribution vanishes, all inertial contributions for the diagonal coframe~\eqref{TetradSS} arise from the contortion tensor, which yields the nonvanishing component
\begin{equation}
K_{12\mu}u^{\mu}= \frac{A_{1}'}{A_{1}A_{2}}.
\end{equation}
The only nonvanishing component of $\Sigma_{ab}$ is
\begin{equation}\label{a2}
\Sigma_{21}=\frac{A_{1}'}{A_{1}A_{2}},
\end{equation}
showing that the inertial frame condition for the diagonal coframe~\eqref{TetradSS} reduces to a constraint on $A_{1}$ alone. Specifically, this particular tetrad represents an inertial comoving frame only if $A_{1}=\text{constant}$, a very restrictive condition that would confine the geometry to the subclass of static, spherically symmetric spacetimes with a spatially homogeneous lapse function. This restriction on the metric function, however, arises from the assumption to employ the diagonal tetrad~\eqref{TetradSS} rather than from the inertial frame condition itself: other tetrad representatives of the same metric, related to the diagonal one by local Lorentz transformations, can satisfy the inertial frame condition without imposing $A_{1}=\text{constant}$.

We now investigate the requirements under which the local Lorentz transformations available in static and spherically symmetric teleparallel geometries can be used to construct inertial comoving frames. In this symmetry class, the available Lorentz transformations consist of a radial boost and a rotation about the radial direction. In the following, we examine each transformation separately. We begin with the boost, which addresses the restriction on the lapse function $A_1$, and then turn to the rotation, which constrains the angular structure of the frame.

\subsubsection{Inertial boosted frame} \label{rapiiner}

We begin with the radial boost, which acts on the timelike and radial directions of the diagonal coframe. The boost is implemented by the local Lorentz transformation $B^{a}{}_{b}=B(x)^{a}{}_{b}$ parametrized by a real-valued function $\psi_{1}=\psi_{1}(r)$~\cite{Emtsova:2021ehh,van2024teleparallel},
\begin{equation}\label{boost}
B^{a}{}_{b}=
\begin{pmatrix}
\cosh\psi_{1} & \sinh\psi_{1} & 0 & 0\\
\sinh\psi_{1} & \cosh\psi_{1} & 0 & 0\\
0 & 0 & 1 & 0\\
0 & 0 & 0 & 1
\end{pmatrix}.
\end{equation}
Applying this transformation to the coframe~\eqref{TetradSS} yields the boosted coframe $\bar{h}^{a}{}_{\mu}=B^{a}{}_{b}h^{b}{}_{\mu}$, whose explicit form is given by
\begin{equation}\label{tetrad3}
\bar h^{a}{}_{\mu}=
\begin{pmatrix}
A_{1}\cosh\psi_{1} & A_{2}\sinh\psi_{1} & 0 & 0\\[1.2ex]
A_{1}\sinh\psi_{1} & A_{2}\cosh\psi_{1} & 0 & 0\\[1.2ex]
0 & 0 & r & 0\\[1.2ex]
0 & 0 & 0 & r\sin\theta
\end{pmatrix}.
\end{equation}
The spin connection~\eqref{SpinV} transforms under the boost \eqref{boost} as described in~\eqref{trans}, yielding the nonvanishing components
\begin{equation} \label{spinv3}
\begin{gathered}
\bar{\omega}_{41\theta} = \bar{\omega}_{13\phi}/\sin\theta = \sin\chi \sinh(\psi-\psi_{1}),\quad
\bar{\omega}_{13\theta} = \bar{\omega}_{14\phi}/\sin\theta = \cos\chi \sinh(\psi-\psi_{1}), \\[1ex]
\bar{\omega}_{42\theta} = \bar{\omega}_{23\phi}/\sin\theta = \sin\chi \cosh(\psi-\psi_{1}), \quad
\bar{\omega}_{23\theta} = \bar{\omega}_{24\phi}/\sin\theta = \cos\chi \cosh(\psi-\psi_{1}), \\[1ex]
\bar{\omega}_{21r} = \psi'-\psi'_{1},
\quad
\bar{\omega}_{43r} = \chi', \quad
\bar{\omega}_{43\phi} = \cos\theta.
\end{gathered}
\end{equation}
For this coframe~\eqref{tetrad3} to be comoving, it must be adapted to a congruence of observers. We therefore define the associated four-velocity as the timelike component of the boosted frame,
\begin{equation}\label{fourv3}
\bar{u}^{\mu}\equiv\bar{h}_{1}{}^{\mu}=\left(\frac{\cosh\psi_{1}}{A_{1}},-\frac{\sinh\psi_{1}}{A_{2}},0,0\right).
\end{equation}
We now determine the conditions under which this coframe defines an inertial comoving frame by evaluating the inertial frame condition~\eqref{inertial}. The inertial effects are encoded in both the spin connection,
\begin{equation}
\bar{\omega}_{12\mu}\bar{u}^{\mu}= \frac{(\psi'-\psi'_{1})\sinh\psi_{1}}{A_{2}},\qquad
\bar{\omega}_{34\mu}\bar{u}^{\mu}= \frac{\chi'\sinh\psi_{1}}{A_{2}},
\end{equation}
and the contortion tensor,
\begin{equation}
\bar{K}_{12\mu}\bar{u}^{\mu}=\frac{A'_{1}\cosh\psi_{1}+A_{1}\psi'\sinh\psi_{1}}{A_{1}A_{2}},
\qquad
\bar{K}_{34\mu}\bar{u}^{\mu}= \frac{\chi'\sinh\psi_{1}}{A_{2}}.
\end{equation}
Part of these contributions cancel between the spin connection and the contortion tensor in $\Sigma_{ab}$. This partial cancellation illustrates concretely that the contortion tensor cannot, in general, be interpreted as purely gravitational, as is sometimes claimed in the literature~\cite{aldrovandi2012teleparallel,maluf2013teleparallel,aldrovandi2008inertia}; rather, it generically contains both inertial and gravitational contributions~\cite{krvsvsak2024teleparallel}. After this cancellation, the remaining contribution to $\Sigma_{ab}$ is
\begin{equation}
\bar{\Sigma}_{21}=\frac{[A_{1}\cosh\psi_{1}]'}{A_{1}A_{2}}.
\end{equation}
For the frame to be inertial, this expression must vanish, requiring
\[
A_{1}\cosh\psi_{1} = \ell,
\]
for some nonzero constant $\ell \in \mathbb{R}$ satisfying $|\ell| \geq |A_{1}|$, since $\psi_{1}$ is real valued. This condition is equivalent to
\begin{equation}\label{psi0A}
\psi_{1}=\pm\cosh^{-1}\left[\frac{\ell}{A_{1}}\right],
\end{equation}
which is the constraint imposed on the non-diagonal frame~\eqref{tetrad3} by the inertial frame condition. For a generic choice of the boost parameter ($\psi_{1}$), this constraint is not satisfied, and the frame~\eqref{tetrad3} is therefore not inertial.

The two signs in~\eqref{psi0A} correspond to two physically distinct branches of the boost related by a reflection of the radial direction. In what follows, we restrict our attention to the positive branch for simplicity. The constant $\ell$ can be absorbed through a rescaling of the time coordinate, $dt' = \ell\, dt$, together with $A_1' = A_1/\ell$. Hence, without loss of generality, we set $\ell = 1$ in what follows. Substituting~\eqref{psi0A} into~\eqref{tetrad3}, we obtain the inertial comoving coframe in the explicit form
\begin{equation}\label{tebar}
\bar h^{a}{}_{\mu}=
\begin{pmatrix}
1 & \dfrac{A_{2}}{A_{1}}\sqrt{1-A_{1}^{2}}& 0 & 0\\[1.2ex]
\sqrt{1-A_{1}^{2}} & \dfrac{ A_{2}}{A_{1}} & 0 & 0\\[1.2ex]
0 & 0 & r & 0\\[1.2ex]
0 & 0 & 0 & r\sin\theta
\end{pmatrix}.
\end{equation}
The associated spin connection has nonvanishing components
\begin{equation}\label{spinbar}
\begin{gathered}
\bar{\omega}_{41\theta} = \bar{\omega}_{13\phi}/\sin\theta = \sin\chi \left(\frac{\sinh\psi}{A_{1}}- \sqrt{ A_{1}^{-2}-1}\, \cosh\psi\right),\\[1ex]
\bar{\omega}_{13\theta} = \bar{\omega}_{14\phi}/\sin\theta = \cos\chi \left(\frac{\sinh\psi}{A_{1}}-\sqrt{A_{1}^{-2}-1}\,\cosh\psi\right), \\[1ex]
\bar{\omega}_{42\theta} = \bar{\omega}_{23\phi}/\sin\theta = \sin\chi \left(\frac{\cosh\psi}{A_{1}}- \sqrt{A_{1}^{-2}-1}\,\sinh\psi\right), \\[1ex]
\bar{\omega}_{23\theta} = \bar{\omega}_{24\phi}/\sin\theta = \cos\chi \left(\frac{\cosh\psi}{A_{1}}- \sqrt{ A_{1}^{-2}-1} \,\sinh\psi\right), \\[1ex]
\bar{\omega}_{21r} = \psi'+\frac{A'_{1}}{A_{1}\sqrt{1-A_{1}^{2}}},
\quad
\bar{\omega}_{43r} = \chi', \quad
\bar{\omega}_{43\phi} = \cos\theta.
\end{gathered}
\end{equation}
This computation leaves $\psi$ and $\chi$ unconstrained, so the Lorentz sector of the teleparallel geometry remains unrestricted. 

A natural special case arises when the boost parameter $\psi_{1}$ coincides with the Lorentz sector function $\psi$ appearing in the original spin connection, i.e., by setting $\psi_{1}=\psi$. Now if we also want to impose the inertial frame condition, the Lorentz sector function is fixed 
\begin{equation}\label{psiIFC}
\psi=\cosh^{-1}\left[\frac{1}{A_{1}}\right].
\end{equation}
The tetrad~\eqref{tebar} remains unchanged, while the associated spin connection simplifies considerably to
\begin{equation} \label{spinbar2}
\begin{gathered}
\bar{\omega}_{42\theta} = \bar{\omega}_{23\phi}/\sin\theta = \sin\chi, \quad
\bar{\omega}_{23\theta} = \bar{\omega}_{24\phi}/\sin\theta = \cos\chi, \\[1ex]
\bar{\omega}_{43r} = \chi', \quad
\bar{\omega}_{43\phi} = \cos\theta.
\end{gathered}
\end{equation}
The pair~\eqref{tebar} and~\eqref{spinbar2} therefore describes an inertial comoving frame for a static, spherically symmetric teleparallel geometry in which the Lorentz sector function $\psi$ is fixed to the form given by~\eqref{psiIFC}.

A radial boost \eqref{boost} with boost parameter \eqref{psi0A} or \eqref{psiIFC} when applied to~\eqref{TetradSS} yields an inertial comoving frame for any spatially inhomogeneous lapse function (i.e.\ $A_{1}$ need not be constant). Although not required for the construction of an inertial comoving frame, the analysis of the constraints arising from the inertial frame condition after a rotation about the radial direction is relevant, as this corresponds to the other residual Lorentz transformation available.

\subsubsection{Inertial rotated frame}  \label{rotaframe}

We now consider the rotation about the radial direction, which acts on the angular components of the boosted coframe. The rotation is implemented by the local Lorentz transformation $R^{a}{}_{b}=R(x)^{a}{}_{b}$ parametrized by a real-valued function $\chi_{1}=\chi_{1}(r)$~\cite{van2024teleparallel},
\begin{equation}\label{rota}
R^{a}{}_{b}=
\begin{pmatrix}
1 & 0 & 0 & 0\\
0 & 1 & 0 & 0\\
0 & 0 & \cos\chi_{1} & -\sin\chi_{1}\\
0 & 0 & \sin\chi_{1} & \cos\chi_{1}
\end{pmatrix}.
\end{equation}
Applying this transformation to the coframe~\eqref{tebar} yields the rotated coframe $\hat{h}^{a}{}_{\mu}=R^{a}{}_{b}\bar{h}^{b}{}_{\mu}$,
\begin{equation}\label{tehat}
\hat{h}^{a}{}_{\mu}=
\begin{pmatrix}
1 & \dfrac{A_{2}}{A_{1}}\sqrt{1-A_{1}^{2}}& 0 & 0\\[1.2ex]
\sqrt{1-A_{1}^{2}} & \dfrac{A_{2}}{A_{1}} & 0 & 0\\[1.2ex]
0 & 0 & r \cos\chi_{1} & -r\sin\chi_{1}\sin\theta\\[1.2ex]
0 & 0 & r\sin\chi_{1} & r\cos\chi_{1}\sin\theta
\end{pmatrix}.
\end{equation}
The spin connection~\eqref{spinbar} transforms under the rotation \eqref{rota} as described in~\eqref{trans}, yielding the nonvanishing components
\begin{equation}\label{spinhat}
\begin{gathered}
\hat{\omega}_{41\theta} = \hat{\omega}_{13\phi}/\sin\theta = \sin(\chi-\chi_{1}) \left(\frac{\sinh\psi}{A_{1}}-\sqrt{A_{1}^{-2}-1}\,\cosh\psi\right),\\[1ex]
\hat{\omega}_{13\theta} = \hat{\omega}_{14\phi}/\sin\theta = \cos(\chi-\chi_{1}) \left(\frac{\sinh\psi}{A_{1}}-\sqrt{A_{1}^{-2}-1}\,\cosh\psi\right), \\[1ex]
\hat{\omega}_{42\theta} = \hat{\omega}_{23\phi}/\sin\theta = \sin(\chi-\chi_{1}) \left(\frac{\cosh\psi}{A_{1}}-\sqrt{A_{1}^{-2}-1}\,\sinh\psi\right), \\[1ex]
\hat{\omega}_{23\theta} = \hat{\omega}_{24\phi}/\sin\theta = \cos(\chi-\chi_{1}) \left(\frac{\cosh\psi}{A_{1}}-\sqrt{A_{1}^{-2}-1} \, \sinh\psi\right), \\[1ex]
\hat{\omega}_{21r} = \psi'+\frac{A'_{1}}{A_{1}\sqrt{1-A_{1}^{2}}},
\quad
\hat{\omega}_{43r} = \chi'-\chi'_{1}, \quad
\hat{\omega}_{43\phi} = \cos\theta.
\end{gathered}
\end{equation}
Since the timelike component of the rotated frame coincides with that of~\eqref{tebar}, the associated four-velocity is again given by~\eqref{fourv3}, that is, $\hat{u}^{\mu}=\bar{u}^{\mu}$. Evaluating the inertial frame condition~\eqref{inertial}, we find that inertial effects are encoded in both the spin connection,
\begin{equation}
\hat{\omega}_{12\mu}\hat{u}^{\mu}= \frac{A'_{1}+A_{1}\psi'\sqrt{1-A_{1}^{2}}}{A_{1}^{2}A_{2}},\qquad
\hat{\omega}_{34\mu}\hat{u}^{\mu}= \frac{(\chi'-\chi'_{1})\sqrt{1-A_{1}^{2}}}{A_{1}A_{2}},
\end{equation}
and the contortion tensor,
\begin{equation}
\hat{K}_{12\mu}\hat{u}^{\mu}=\frac{A'_{1}+A_{1}\psi'\sqrt{1-A_{1}^{2}}}{A_{1}^{2}A_{2}},
\qquad
\hat{K}_{34\mu}\hat{u}^{\mu}= \frac{\chi'\sqrt{1-A_{1}^{2}}}{A_{1}A_{2}}.
\end{equation}
Part of these contributions cancel between the spin connection and the contortion tensor in $\Sigma_{ab}$. After the cancellation, the remaining contribution is
\begin{equation}
\hat{\Sigma}_{43}=\frac{\chi'_{1}\sqrt{1-A_{1}^{2}}}{A_{1}A_{2}}.
\end{equation}
For the frame to be inertial, this expression must vanish, which requires $\chi_{1}$ to be constant,
\begin{equation}\label{chi1v}
\chi_{1}=\chi_{0},\qquad \chi_{0}\in\mathbb{R}.
\end{equation}
This computation leaves $\psi$ and $\chi$ unconstrained, so the Lorentz sector of the teleparallel geometry again remains unrestricted.

As in the boosted case, we consider the special situation in which the boost parameter $\psi_{1}$ and the rotation parameter $\chi_{1}$ coincide with the Lorentz sector functions $\psi$ and $\chi$ appearing in the original spin connection, i.e., $\psi_{1}=\psi$ and $\chi_{1}=\chi$. Imposing the inertial frame condition then yields constraints on the Lorentz sector functions. As before, $\psi$ is required to be given by \eqref{psiIFC}, while $\chi$ must satisfy
\begin{equation}\label{chiIFC}
\chi=\chi_{0} .
\end{equation}
The tetrad~\eqref{tehat} is unchanged, while the associated spin connection simplifies to
\begin{equation}\label{spinhat2}
\begin{gathered}
\hat{\omega}_{23\theta} = \hat{\omega}_{24\phi}/\sin\theta = 1, \quad
\hat{\omega}_{43\phi} = \cos\theta.
\end{gathered}
\end{equation}
The pair~\eqref{tehat} and \eqref{spinhat2} therefore describes an inertial comoving frame for a static, spherically symmetric teleparallel geometry with a Lorentz sector more restricted than in the simply boosted case~\eqref{tebar} and \eqref{spinbar2}.

A constant rotation about the radial direction therefore maps inertial comoving frames to inertial comoving frames. This observation admits a natural generalization. From condition~\eqref{conG}, the local Lorentz transformations that preserve the inertial character of the frame satisfy a stronger requirement than mere constancy along the observer's worldline, since the transformation may alter the observer congruence. In the static and spherically symmetric case, the most general local Lorentz transformation compatible with the symmetry takes the form~\cite{van2024teleparallel}
\begin{equation}\label{LTg}
\Lambda^{a}{}_{b} =
\begin{pmatrix}
\ch_{\psi_{2}} & \sh_{\psi_{2}} & 0 & 0 \\[6pt]
\co_{\theta}\sh_{\psi_{2}} &
\co_{\theta}\ch_{\psi_{2}} &
\si_{\theta}\co_{\chi_{2}} &
-\si_{\theta}\si_{\chi_{2}} \\[6pt]
- \si_{\theta} \co_{\phi}\sh_{\psi_{2}} &
- \si_{\theta}\co_{\phi}\ch_{\psi_{2}} &
\co_{\theta}\co_{\phi}\co_{\chi_{2}} - \si_{\phi}\si_{\chi_{2}} &
- \co_{\theta}\co_{\phi}\si_{\chi_{2}} - \si_{\phi}\co_{\chi_{2}} \\[6pt]
- \si_{\theta}\si_{\phi}\sh_{\psi_{2}} &
- \si_{\theta}\si_{\phi}\ch_{\psi_{2}} &
\co_{\phi}\si_{\chi_{2}} + \co_{\theta}\si_{\phi}\co_{\chi_{2}} &
\co_{\phi}\co_{\chi_{2}} - \co_{\theta}\si_{\phi}\si_{\chi_{2}}
\end{pmatrix},
\end{equation}
where $\psi_{2}=\psi_{2}(r)$ and $\chi_{2}=\chi_{2}(r)$ are real-valued functions of the radial coordinate, and we use the shorthand
\begin{equation}
\si_{y}\equiv\sin y, \quad
\co_{y}\equiv\cos y, \quad
\sh_{y}\equiv\sinh y, \quad
\ch_{y}\equiv\cosh y.
\end{equation}
The angles $\theta$ and $\phi$ appearing in~\eqref{LTg} encode the angular dependence required by spherical symmetry and do not constitute independent local Lorentz degrees of freedom, in contrast to $\psi_{2}$ and $\chi_{2}$. Substituting~\eqref{LTg} into~\eqref{conG}, one finds that the inertial character of the frame is preserved if and only if
\begin{equation}
\psi_{2}=0, \qquad \chi_{2}=\chi_{0}.
\end{equation}
The admissible transformations therefore reduce to constant rotations about the radial direction (parame-trized by $\chi_{0}$), while no nontrivial boost preserves the inertial character: the boost parameter must vanish identically. This reflects the geometric content of condition~\eqref{conG}: a nontrivial $\psi_{2}$ would alter the observer congruence, and the inertial frame condition could no longer be preserved.

\subsubsection{Inertial proper frame}

As established in Section~\ref{rproper}, the proper frame alone is insufficient to eliminate inertial effects; the inertial frame condition~\eqref{inertial} must be separately imposed. We now outline the procedure for constructing an inertial proper frame in static and spherically symmetric teleparallel geometries.

We begin by transforming the diagonal coframe~\eqref{TetradSS} into the proper frame using the Lorentz transformation~\eqref{LTg} with $\psi_{2}=\psi$ and $\chi_{2}=\chi$, which maps the tetrad~\eqref{TetradSS} and spin connection~\eqref{SpinV} to
\begin{equation}
\tilde{h}^{a}{}_{\mu}=\Lambda^{a}{}_{b}h^{b}{}_{\mu}, \qquad \tilde{\omega}^{a}{}_{b\mu}=0.
\end{equation}
The resulting proper frame tetrad takes the explicit form~\cite{van2024teleparallel}
\begin{equation}\label{propert}
\tilde{h}^{a}{}_{\mu} =
\begin{pmatrix}
A_{1}\ch_{\psi} & A_{2}\sh_{\psi} & 0 & 0 \\[6pt]
A_{1}\co_{\theta}\sh_{\psi} &
A_{2}\co_{\theta}\ch_{\psi} &
r\si_{\theta}\co_{\chi} &
-r\si_{\theta}^{2}\si_{\chi} \\[6pt]
-A_{1}\si_{\theta}\co_{\phi}\sh_{\psi} &
-A_{2}\si_{\theta}\co_{\phi}\ch_{\psi} &
r(\co_{\theta}\co_{\phi}\co_{\chi} - \si_{\phi}\si_{\chi}) &
-r\si_{\theta}(\co_{\theta}\co_{\phi}\si_{\chi} + \si_{\phi}\co_{\chi}) \\[6pt]
-A_{1}\si_{\theta}\si_{\phi}\sh_{\psi} &
-A_{2}\si_{\theta}\si_{\phi}\ch_{\psi} &
r(\co_{\phi}\si_{\chi} + \co_{\theta}\si_{\phi}\co_{\chi}) &
r\si_{\theta}(\co_{\phi}\co_{\chi} - \co_{\theta}\si_{\phi}\si_{\chi})
\end{pmatrix}.
\end{equation}
The four-velocity associated with this coframe is the timelike component $\tilde{u}^{\mu}\equiv\tilde{h}_{1}{}^{\mu}$, which coincides with~\eqref{fourv3}. Since the spin connection vanishes identically in the proper frame, the inertial contributions to condition~\eqref{inertial} arise entirely from the contortion tensor. Evaluating~\eqref{inertial} yields
\begin{equation}
\tilde{\Sigma}_{ab} =
\begin{pmatrix}
0 &
-\dfrac{\co_{\theta}[\ch_{\psi}A_{1}]'}{A_{1}A_{2}} &
\dfrac{\si_{\theta}\co_{\phi}[\ch_{\psi}A_{1}]'}{A_{1}A_{2}} &
\dfrac{\si_{\theta}\si_{\phi}[\ch_{\psi}A_{1}]'}{A_{1}A_{2}}
\\[10pt]
\dfrac{\co_{\theta}[\ch_{\psi}A_{1}]'}{A_{1}A_{2}} &
0 &
\dfrac{\si_{\theta}\si_{\phi}\sh_{\psi}\chi'}{A_{2}} &
-\dfrac{\si_{\theta}\co_{\phi}\sh_{\psi}\chi'}{A_{2}}
\\[10pt]
-\dfrac{\si_{\theta}\co_{\phi}[\ch_{\psi}A_{1}]'}{A_{1}A_{2}} &
-\dfrac{\si_{\theta}\si_{\phi}\sh_{\psi}\chi'}{A_{2}} &
0 &
-\dfrac{\co_{\theta}\sh_{\psi}\chi'}{A_{2}}
\\[10pt]
-\dfrac{\si_{\theta}\si_{\phi}[\ch_{\psi}A_{1}]'}{A_{1}A_{2}} &
\dfrac{\si_{\theta}\co_{\phi}\sh_{\psi}\chi'}{A_{2}} &
\dfrac{\co_{\theta}\sh_{\psi}\chi'}{A_{2}} &
0
\end{pmatrix}.
\end{equation}
For the proper frame to be inertial, all components of $\tilde{\Sigma}_{ab}$ must vanish. This requires the Lorentz sector functions $\psi$ and $\chi$ to take the values \eqref{psiIFC} and \eqref{chiIFC}, respectively. Substituting these conditions into~\eqref{propert}, we obtain the inertial proper frame in the explicit form
\begin{equation}\label{propertv}
\tilde{h}^{a}{}_{\mu} =
\begin{pmatrix}
1 & \dfrac{A_{2}}{A_{1}}\sqrt{1-A_{1}^{2}} & 0 & 0 \\[9pt]
\co_{\theta}\sqrt{1-A_{1}^{2}} &
\dfrac{A_{2}}{A_{1}}\co_{\theta} &
r\si_{\theta}\co_{\chi_{0}} &
-r\si_{\theta}^{2}\si_{\chi_{0}} \\[9pt]
-\si_{\theta}\co_{\phi}\sqrt{1-A_{1}^{2}} &
-\dfrac{A_{2}}{A_{1}}\si_{\theta}\co_{\phi} &
r(\co_{\theta}\co_{\phi}\co_{\chi_{0}} - \si_{\phi}\si_{\chi_{0}}) &
-r\si_{\theta}(\co_{\theta}\co_{\phi}\si_{\chi_{0}} + \co_{\chi_{0}}\si_{\phi}) \\[9pt]
-\si_{\theta}\si_{\phi}\sqrt{1-A_{1}^{2}} &
-\dfrac{A_{2}}{A_{1}}\si_{\theta}\si_{\phi} &
r(\co_{\phi}\si_{\chi_{0}} + \co_{\theta}\si_{\phi}\co_{\chi_{0}}) &
r\si_{\theta}(\co_{\phi}\co_{\chi_{0}} - \co_{\theta}\si_{\phi}\si_{\chi_{0}})
\end{pmatrix}.
\end{equation}
The inertial proper frame~\eqref{propertv} is not unique: it depends on the constant value $\chi_{0}$, which labels the members of the equivalence class of inertial proper frames, which are related to one another by constant rotations about the radial direction.

Unlike the inertial comoving frames constructed in the preceding subsections, where $\psi$ and $\chi$ may remain arbitrary, the construction of an inertial proper frame requires fixing both $\psi$ and $\chi$ according to~\eqref{psiIFC} and~\eqref{chiIFC}. The inertial proper frame therefore belongs to the same restricted class as the pair~\eqref{tehat} and \eqref{spinhat2}, in which both $\psi$ and $\chi$ are fixed.

The inertial frames constructed in this section are adapted to free-fall observers in static and spherically symmetric teleparallel geometries. They are valid in any teleparallel theory with minimally coupled matter, since the inertial frame condition~\eqref{inertial} arises from the equation of motion common to such theories. All such frames belong to a single equivalence class related by constant rotations about the radial direction, within which inertial effects are consistently eliminated along the observer's worldline.

\subsection{Torsion scalar invariant}\label{torsca}

The invariance of the torsion scalar under local Lorentz transformations implies that it is insensitive to some Lorentz parameters. This motivates a precise distinction between two classes of Lorentz parameters. We define the \emph{residual Lorentz freedom} as the subgroup of local Lorentz transformations relating physically equivalent representatives of a given teleparallel geometry. This stands in contrast to the Lorentz sector, which distinguishes inequivalent teleparallel geometries sharing the same metric structure: while the Lorentz sector functions enter the torsion scalar, the parameters associated with the residual Lorentz freedom do not.

In the static and spherically symmetric case, the residual Lorentz freedom is parametrized by two real-valued functions of the radial coordinate associated with a radial boost ($\psi_{1}$) and a rotation about the radial direction ($\chi_{1}$). Likewise, the Lorentz sector is parametrized by two real-valued functions associated with a boost ($\psi$) and a rotation ($\chi$). Although the tetrad and spin connection may in general depend on all four functions $(\psi_{1}, \chi_{1}, \psi, \chi)$, the torsion scalar depends only on $(\psi, \chi)$. Explicitly, for the tetrad~\eqref{TetradSS} and spin connection~\eqref{SpinV}, the torsion scalar takes the form
\begin{equation}\label{tor1}
T =
-\frac{2 \left(A_{1} (1 + A_{2}^{2}) + 2 r A_{1}'\right)}{A_{1} A_{2}^{2} r^{2}}
-\frac{4 \cos\chi \cosh\psi \left(A_{1} + r A_{1}'\right)}{A_{1} A_{2} r^{2}}
-\frac{4 [\cos\chi \cosh\psi]'}{A_{2} r} \, .
\end{equation}
This is confirmed by the fact that computing the torsion scalar using the boosted representatives~\eqref{tetrad3} and~\eqref{spinv3}, or the rotated representatives~\eqref{tehat} and~\eqref{spinhat}, yields exactly the same expression~\eqref{tor1}, making the distinction between the two classes of Lorentz parameters explicit.

The dependence of~\eqref{tor1} on $\psi$ and $\chi$ has been interpreted in the literature as reflecting inertial contributions to the torsion scalar~\cite{krvsvsak2015spin,krvsvsak2019teleparallel}. This interpretation should be revisited. As established in the preceding subsection, making a frame comoving and inertial does not require fixing the Lorentz sector functions $\psi$ and $\chi$; indeed, computing the torsion scalar using the inertial comoving representatives~\eqref{tebar} and~\eqref{spinbar}, or~\eqref{tehat} and~\eqref{spinhat}, yields exactly the same expression~\eqref{tor1}. The dependence on $\psi$ and $\chi$ is therefore not an inertial dependence in the dynamical sense of~\eqref{inertial}, but rather a dependence on the Lorentz sector distinguishing inequivalent teleparallel geometries. Inertial contributions and the dependence of the torsion scalar on the Lorentz sector therefore constitute two independent structures in TG.

The distinction between different teleparallel geometries within the same static and spherically symmetric class can be made explicit by fixing the Lorentz sector. Consider the value of $\psi$ given by~\eqref{psiIFC} for constructing an inertial comoving frame as described in Subsection~\ref{rapiiner}. Under this choice for $\psi$, the torsion scalar takes the form
\begin{equation}\label{tor2}
T=-\frac{2\left(A_{1}\left(1+A_{2}^{2}\right)
+ 2r A_{1}'
+ 2A_{2}\,\left(\cos\chi - r \sin\chi \, \chi' \right)\right)}
{A_{1} A_{2}^{2} r^{2}},
\end{equation}
which corresponds to the subclass of static and spherically symmetric teleparallel geometries parametrized by the function $\chi$. Expression~\eqref{tor2} is evidently different from~\eqref{tor1}, as are the corresponding geometries they describe. In particular, the general class contains unrestricted $\psi$ and $\chi$, while the subclass considered here restricts $\psi$ while leaving $\chi$ unrestricted.

A further restriction is obtained by fixing $\chi$ in addition to $\psi$. This can be achieved using the inertial comoving frame constructed in Subsection~\ref{rotaframe}, which requires $\chi$ to be constant according to~\eqref{chiIFC}. The resulting subclass is therefore more restrictive, and the torsion scalar reduces to
\begin{equation}\label{tor3}
T=-\frac{2\left(A_{1}\left(1+A_{2}^{2}\right)
+2r\,A_{1}'
+2A_{2}\,\cos\!\chi\right)}
{A_{1}A_{2}^{2}r^{2}}.
\end{equation}
Note that the same expression~\eqref{tor3} is obtained using the inertial proper frame~\eqref{propertv}, making explicit that the particular representative used to compute the torsion scalar is irrelevant, provided it belongs to the same equivalence class.

The preceding analysis implies that the divergences encountered in the TEGR literature are not a consequence of inertial effects in the sense of condition~\eqref{inertial}; they are genuine features of a subclass of teleparallel geometries within the Schwarzschild family, characterized by specific values of the Lorentz sector functions. This motivates the search for the complementary subclass of teleparallel geometries that, unlike the commonly studied one, is compatible with a regular horizon description. 

\section{Schwarzschild TEGR geometries}\label{SchTEGR}

The Schwarzschild geometry is the unique static, spherically symmetric vacuum solution of the Einstein field equations of GR. It describes the exterior gravitational field of a non-rotating, uncharged, spherically symmetric mass~\cite{wald2010general,hawking2023large}, characterized by
\begin{equation}\label{SchA}
A_{1}=A_{2}^{-1}=\sqrt{1-\frac{r_{h}}{r}},
\end{equation}
where $r_{h}$ is the location of the event horizon. Although the metric components appear singular at $r=r_{h}$ in these coordinates, this singularity is purely coordinate in nature and can be removed by an appropriate change of coordinates~\cite{wald2010general,hawking2023large}. In contrast, curvature scalar invariants such as the Kretschmann scalar diverge at $r=0$, signaling the presence of a genuine physical singularity. The Schwarzschild spacetime is asymptotically flat and reduces to Minkowski spacetime in the limit $r_{h}\to 0$~\cite{wald2010general}. It provides the canonical model of a static black hole in GR and serves as the fundamental reference solution for comparisons with alternative theories of gravity, including teleparallel formulations.

Since TEGR admits the same vacuum solutions as GR, the metric functions are given by~\eqref{SchA}~\cite{aldrovandi2012teleparallel}. However, since the teleparallel geometry is described by a tetrad and an associated spin connection rather than by the metric alone, the solution is not unique in TEGR. The functions $\psi$ and $\chi$ do not enter the TEGR FE~\cite{van2024teleparallel,mcnutt2023frame} and remain undetermined by them: they constitute the Lorentz sector that survives after the TEGR FE and symmetry requirements have been imposed.

In GR this freedom is hidden, since the metric is invariant under local Lorentz transformations and the geometry is uniquely specified by $g_{\mu\nu}$. In TEGR, by contrast, the Lorentz sector is explicit~\cite{weatherall2024general,krvsvsak2019teleparallel}, as is directly visible in the torsion scalar~\eqref{tor1}, which when specialized to the Schwarzschild case~\eqref{SchA} becomes
\begin{dmath}\label{torSch}
T = -\frac{4}{r^{2}}
+ \frac{2 \co_{\chi} \ch_{\psi} \left(r_{h} - 2r\right)}{r^{5/2} \sqrt{r-r_{h}}}
+ \frac{4 \ch_{\psi} \si_{\chi} \sqrt{r-r_{h}}\, \chi'}{r^{3/2}}
- \frac{4 \co_{\chi} \sh_{\psi} \sqrt{r-r_{h}}\, \psi'}{r^{3/2}}.
\end{dmath}
The torsion scalar exhibits a potential divergence at $r=r_{h}$~\cite{obukhov2003metric,coley2025black,lopez2025black}, as is explicit in~\eqref{torSch}. The same potential divergence at $r=r_{h}$ appears in the scalar invariants~\eqref{AVTS}:
\begin{subequations}
\begin{dmath}\label{aSch}
\mathscr{A} =
-\frac{16\si_{\chi}^{2}}{9r^{2}}
+ \frac{16\sqrt{r-r_{h}}\,\ch_{\psi}\,\si_{\chi}\,\chi'}{9r^{3/2}}
+ \frac{4(r_{h}-r)(\chi')^{2}}{9r},
\end{dmath}
\begin{dmath}\label{vSch}
\mathscr{V} =
\frac{4\co_{\chi}^{2}}{r^{2}}
+ \frac{(3r_{h}-4r)^{2}}{4r^{3}(r-r_{h})}
+ \frac{2\co_{\chi}\ch_{\psi}\,(4r-3r_{h})}{\sqrt{r^{5}(r-r_{h})}}
+ \frac{4\co_{\chi}\sh_{\psi}\sqrt{r-r_{h}}\,\psi'}{r^{3/2}}
+ \frac{(r_{h}-r)(\psi')^{2}}{r},
\end{dmath}
\begin{dmath}\label{tSch}
\mathscr{T} =
\frac{1}{r^{2}}
+ \frac{(3r_{h}-2r)^{2}}{4r^{3}(r-r_{h})}
+ \frac{\co_{\chi}\ch_{\psi}\,(2r-3r_{h})}{\sqrt{r^{5}(r-r_{h})}}
+\frac{2\sqrt{r-r_{h}}}{r^{3/2}}
\left(\ch_{\psi}\si_{\chi}\,\chi'
-\co_{\chi}\sh_{\psi}\,\psi'
\right)
\\- \frac{(r_{h}-r)({\chi'}^{2}-{\psi'}^{2})}{r}.
\end{dmath}
\end{subequations}
Although the functions $\psi$ and $\chi$ are not fixed by the TEGR FE, their behaviour at $r=r_{h}$ must also be taken into account. Even in cases where no explicit divergence appears, as in the case of $\mathscr{A}$ in~\eqref{aSch}, which contains no manifest singularity at $r=r_{h}$, particular values of $\chi$ or its derivatives may nevertheless introduce one.  

At first sight, the divergences in~\eqref{torSch}, \eqref{vSch}, and~\eqref{tSch} pose an interpretational challenge, as they suggest that the Schwarzschild TEGR geometries defined by~\eqref{SchA} are singular at both $r=0$ and $r=r_h$. Consequently, the horizon $r=r_h$ would be excluded from the spacetime manifold, obstructing the usual black-hole interpretation. However, given the freedom in the Lorentz sector functions $\psi$ and $\chi$, the divergence at $r=r_h$ need not be a universal feature of Schwarzschild TEGR geometries. Although the generic description appears incompatible with a regular horizon, $\psi$ and $\chi$ remain completely arbitrary functions, a property that can be exploited to identify values compatible with a regular horizon. This raises the question of how the Lorentz sector functions should be selected and which criteria can be used to identify values compatible with a regular horizon.

For example, it could be asserted that any viable gravitational theory must admit a regime in which all gravitational effects vanish and the description reduces to that of SR. For Schwarzschild TEGR geometries, this regime is reached by taking $r_{h}\to 0$, which recovers flat spacetime~\cite{krvsvsak2015spin,lucas2009regularizing}. We therefore require that all torsion components constructed from the tetrad~\eqref{TetradSS} and the spin connection~\eqref{SpinV} with~\eqref{SchA} vanish in this limit. Imposing this condition on the torsion constraints the limiting behaviour of $\chi$ and $\psi$: they must approach constant values $\chi\to\chi_{0}$ and $\psi\to\psi_{0}$, with $\chi_{0},\psi_{0}\in\mathbb{R}$, satisfying
\begin{equation}
\cos\chi_{0}\sinh\psi_{0}=0, \qquad \cos\chi_{0}\cosh\psi_{0}=-1.
\end{equation}
These conditions are simultaneously satisfied only if
\begin{equation}\label{limitxy}
\lim_{r_{h}\to 0}\chi(r)=(2n+1)\pi, \qquad \lim_{r_{h}\to 0}\psi(r)=0,
\end{equation}
where $n\in\mathbb{Z}$. The conditions in~\eqref{limitxy} do not fix $\chi$ and $\psi$ \emph{a priori} in the general geometric setup; rather, they constrain only the asymptotic behaviour of these functions. Consequently, any pair of functions satisfying~\eqref{limitxy} reproduces the correct gravity-free limit. As a simple example, consider
\begin{equation}\label{pchi}
\psi(r) = r_{h}\,Y(r),
\quad\text{and}\quad
\chi(r) = (2n+1)\pi + r_{h}\,X(r),
\end{equation}
where $Y(r)$ and $X(r)$ are real-valued differentiable functions. The expressions in~\eqref{pchi} reduce to $\psi=0$ and $\chi=(2n+1)\pi$ as $r_{h}\to 0$, in agreement with~\eqref{limitxy}. One may therefore set $\chi=(2n+1)\pi$ and $\psi=0$, and evaluate the resulting torsion scalar invariants. However, the divergences at the horizon persist, demonstrating that recovering the correct gravity-free limit is not sufficient to guarantee horizon regularity. Additional criteria are therefore needed. In what follows, we present three complementary approaches: (i) the inertial frame condition, (ii) horizon-penetrating coordinates, and (iii) the horizon regularity criterion.

\subsection{Inertial frame condition}

As established in Section~\ref{torsca}, the torsion scalar invariants are independent of inertial effects in the dynamical sense of condition~\eqref{inertial}, so the inertial frame condition has no a priori connection to horizon regularity. Nevertheless, because the TEGR FE leave $\psi$ and $\chi$ undetermined, the Lorentz sector of the Schwarzschild TEGR geometry is not fixed by the dynamics. The inertial frame condition~\eqref{inertial} therefore provides a physically motivated approach to selecting specific values of $\psi$ and $\chi$.

Let us consider the inertial coframe~\eqref{tetrad3} in the special case where the boost parameter $\psi_{1}$ is identified with the Lorentz sector function $\psi$, that is, $\psi_{1}=\psi$. Imposing the inertial frame condition then fixes $\psi$ to the value given in~\eqref{psiIFC}, which, upon specializing to the Schwarzschild geometry~\eqref{SchA}, becomes
\begin{equation}\label{psi0ASch}
\psi=\cosh^{-1}\left[\frac{\sqrt{r}}{\sqrt{r-r_{h}}}\right].
\end{equation}
This choice of $\psi$ also satisfies the asymptotic condition~\eqref{limitxy}. The resulting inertial coframe for the Schwarzschild TEGR geometry is
\begin{equation}\label{te3Sch}
\bar{h}^{a}{}_{\mu}=
\begin{pmatrix}
1 & \dfrac{\sqrt{r_{h}r}}{r-r_{h}} & 0 & 0\\[1.2ex]
\sqrt{\dfrac{r_{h}}{r}} & \dfrac{r}{r-r_{h}} & 0 & 0\\[1.2ex]
0 & 0 & r & 0\\[1.2ex]
0 & 0 & 0 & r\sin\theta
\end{pmatrix}.
\end{equation}
The associated spin connection retains the form~\eqref{spinbar2}. Since it is independent of $A_{1}$ and $\psi$, it is unaffected by the specialization to the Schwarzschild metric functions.

The coframe~\eqref{te3Sch}, together with the spin connection~\eqref{spinbar2}, defines a subclass of Schwarzschild TEGR geometries selected by the inertial frame condition. To determine whether this class is compatible with a regular horizon description, we substitute the value of $\psi$ from~\eqref{psi0ASch} into the torsion scalar~\eqref{torSch}, yielding
\begin{equation}\label{torp}
T=-\frac{4\left(1+\cos\chi-r\sin\chi\,\chi'\right)}{r^{2}}.
\end{equation}
The divergence previously encountered at $r=r_h$ in~\eqref{torSch} is therefore absent. The same is true for the scalar invariants~\eqref{aSch}, \eqref{vSch}, and~\eqref{tSch}, which, after substituting~\eqref{psi0ASch}, take the form
\begin{subequations}\label{avtp}
\begin{equation}
\mathscr{A}=\frac{4\left(r\,\chi'\left(4\sin\chi+(r_{h}-r)\chi'\right)-4\sin^{2}\chi\right)}{9r^{2}},
\end{equation}
\begin{equation}
\mathscr{V}=\frac{64r\cos^{4}(\chi/2)-9r_{h}}{4r^{3}},
\end{equation}
\begin{equation}
\mathscr{T}=\frac{8(1+\cos\chi)r+4r^{2}\chi'\left(2\sin\chi+(r-r_{h})\chi'\right)-9r_{h}}{4r^{3}}.
\end{equation}
\end{subequations}
The function $\chi$ remains otherwise unrestricted, but must satisfy~\eqref{limitxy} and remain finite at $r\geq r_{h}$ together with its derivative,
\begin{equation}\label{chifinite}
\chi(r),\chi'(r)\in\mathbb{R}.
\end{equation}

As a further illustration, we consider the inertial proper coframe~\eqref{propert}, which already satisfies the inertial frame condition. Specializing~\eqref{propert} to the Schwarzschild geometry~\eqref{SchA} and imposing~\eqref{limitxy}, we find that $\psi$ is given by~\eqref{psi0ASch}, while $\chi$ reduces to
\begin{equation}\label{chiSch}
\chi=(2n+1)\pi.
\end{equation}
With the Lorentz sector functions fixed by~\eqref{psi0ASch} and~\eqref{chiSch}, the inertial proper coframe~\eqref{propert} takes the form
\begin{equation}\label{propertsch}
\tilde{h}^{a}{}_{\mu}=
\begin{pmatrix}
1 &
\dfrac{\sqrt{r_{h}r}}{r-r_{h}} &
0 & 0 \\[10pt]
\co_{\theta}\sqrt{\dfrac{r_{h}}{r}} &
\dfrac{\co_{\theta}r}{r-r_{h}} &
-\si_{\theta}r &
0 \\[10pt]
-\co_{\phi}\si_{\theta}\sqrt{\dfrac{r_{h}}{r}} &
\dfrac{\co_{\phi}\si_{\theta}r}{r_{h}-r} &
-r\co_{\theta}\co_{\phi} &
\si_{\theta}\si_{\phi}r \\[10pt]
-\si_{\theta}\si_{\phi}\sqrt{\dfrac{r_{h}}{r}} &
\dfrac{\si_{\theta}\si_{\phi}r}{r_{h}-r} &
-\co_{\theta}\si_{\phi}r &
-\co_{\phi}\si_{\theta}r
\end{pmatrix}.
\end{equation}
This coframe corresponds to a Schwarzschild TEGR geometry obtained by imposing a further restriction on the Lorentz sector within the subclass selected by the inertial frame condition. The corresponding torsion scalar invariants are obtained from~\eqref{torp} and \eqref{avtp} by imposing~\eqref{chiSch} and are given by
\begin{equation}\label{T0}
T=0\,, \quad \mathscr{A}=0\,, \quad \mathscr{V}=\mathscr{T}=-\frac{9r_{h}}{4r^{3}}\,.
\end{equation}
This result parallels the GR case, where the Schwarzschild solution satisfies $R=0$. The divergences previously reported at $r=r_{h}$~\cite{obukhov2003metric,coley2025black,krvsvsak2015spin} arise from choices of the Lorentz sector that are incompatible with a regular horizon description and are absent throughout the present subclass of Schwarzschild TEGR geometries. 

As further evidence that suitable choices of the Lorentz sector functions $\psi$ and $\chi$ yield a regular horizon description, we compute several higher-order contortion scalar invariants,
\vspace{-1.5em}
\begin{subequations}
\begin{center}
\begin{minipage}[t]{0.38\textwidth}
\vspace{0pt}
\begin{flalign}
&
\nabla_{\sigma}K_{\alpha\beta\rho}
\nabla^{\rho}K^{\alpha\beta\sigma}
=
-\frac{255\,r_{h}}{8\,r^{5}} ,
&
\\[1ex]
&
K^{\alpha\beta\sigma}
K_{\sigma}{}^{\rho\mu}
\nabla_{\rho}K_{\alpha\beta\mu}
=
\frac{45\,r_{h}^{2}}{8\,r^{6}} ,
&
\\[1ex]
&
K_{\alpha\beta}{}^{\rho}
K^{\alpha\beta\sigma}
K_{\sigma}{}^{\mu\nu}
K_{\rho\nu\mu}
=
0 ,
&
\end{flalign}
\end{minipage}
\hspace{0.05\textwidth}
\begin{minipage}[t]{0.45\textwidth}
\vspace{0pt}
\begin{flalign}
&
\nabla^{\alpha}K^{\beta\sigma\rho}
\nabla_{\alpha}K_{\beta\sigma\rho}
=
\frac{9\,r_{h}(9\,r_{h}-25\,r)}{8\,r^{6}} ,
&
\\[1ex]
&
K^{\alpha\beta\sigma}
K_{\sigma}{}^{\rho\mu}
\nabla_{\mu}K_{\alpha\beta\rho}
=
0 ,
&
\\[1ex]
&
K_{\alpha}{}^{\rho}{}_{\sigma}
K^{\alpha\beta\sigma}
K_{\beta}{}^{\mu\nu}
K_{\rho\mu\nu}
=
\frac{57\,r_{h}^{2}}{8\,r^{6}} .
&
\end{flalign}
\end{minipage}
\end{center}
\end{subequations}

These results demonstrate that horizon regularity is a property of the Lorentz sector values themselves, rather than of the inertial frame condition that motivated their selection. Equivalently, since the torsion scalar invariants depend only on $\psi$ and $\chi$ and are insensitive to the residual Lorentz freedom, every Schwarzschild TEGR geometry belonging to the same equivalence subclass exhibits the same regular horizon description.

\subsection{Horizon-penetrating coordinates}

We now consider a complementary approach to selecting values of the Lorentz sector functions compatible with a regular horizon description. Rather than imposing the inertial frame condition, this approach is based on the assumption that the Schwarzschild TEGR geometry admits an analytic extension across the horizon. Consequently, we require the tetrad and spin connection to remain regular at $r=r_h$, or equivalently, when working in the proper-frame gauge, that the tetrad itself remain regular there. We start by introducing horizon-penetrating coordinates and subsequently determine the values of $\psi$ and $\chi$ required for a regular extension through the horizon.

A closely analogous situation occurs in GR. The Schwarzschild metric written in Schwarzschild coordinates is singular at $r=r_h$. However, this singularity is purely a coordinate artifact: there exist horizon-penetrating coordinate systems in which the metric remains regular at the horizon~\cite{misner1973gravitation}. Equivalently, one may perform a coordinate transformation such that the associated proper frame remains regular at $r=r_h$.

It has been claimed that, in the proper frame, the causal structure forces the frame to become singular at the horizon, since the normalized timelike tetrad component cannot smoothly approach a null direction there~\cite{golovnev2024static}. A related claim is that the torsion scalar and its associated invariants necessarily diverge at the horizon for Schwarzschild solutions~\cite{obukhov2003metric,coley2025black}. The latter conclusion, however, holds only for Lorentz sector values that are incompatible with a regular horizon description. Indeed, the frame~\eqref{propertsch}, although singular at the horizon, yields torsion scalar invariants that remain finite at $r=r_h$. This shows that the regularity of the torsion scalar invariants is not contingent on the regularity of the frame itself. We therefore revisit the former claim and investigate whether a proper frame can be constructed that remains regular at the horizon.

To this end, we begin by considering coordinate transformations that render the Schwarzschild geometry regular at $r=r_h$. In GR, these transformations give rise to horizon-penetrating coordinate systems. A well-known example is the Eddington--Finkelstein coordinates (EFC), obtained from Schwarzschild coordinates through an appropriate coordinate transformation. To introduce them, we first define the tortoise coordinate
\begin{equation}
r_{*} = r + r_{h}\ln\left|\frac{r}{r_{h}} - 1\right|,
\end{equation}
which modifies the radial structure of the metric so that radial null rays satisfy $t \pm r_{*}=\text{constant}$. The advanced and retarded null coordinates are then defined by
\begin{equation}\label{udef}
u = t + \varepsilon r_{*},
\qquad
du = dt + \varepsilon\,\frac{r}{r - r_{h}}\,dr,
\end{equation}
where $\varepsilon = \pm 1$ distinguishes between the advanced ($+$) and retarded ($-$) cases. Under the coordinate transformation $x^{\mu}=(t,r,\theta,\phi)\to y^{\mu}=(u,r,\theta,\phi)$, the components of the boosted coframe~\eqref{tetrad3}, specialized to the Schwarzschild case~\eqref{SchA}, transform according to
\begin{equation}\label{fef}
\bar{H}^{a}{}_{\mu}=\frac{\partial x^{\nu}}{\partial y^{\mu}}\bar{h}^{a}{}_{\nu}=
\begin{pmatrix}
\sqrt{\dfrac{r-r_{h}}{r}}\cosh\psi &
\sqrt{\dfrac{r}{r-r_{h}}}\left(\sinh\psi-\varepsilon\cosh\psi\right) & 0 & 0 \\[8pt]
\sqrt{\dfrac{r-r_{h}}{r}}\sinh\psi &
\sqrt{\dfrac{r}{r-r_{h}}}\left(\cosh\psi-\varepsilon\sinh\psi\right) & 0 & 0 \\[8pt]
0 & 0 & r & 0 \\[8pt]
0 & 0 & 0 & r\sin\theta
\end{pmatrix}.
\end{equation}
The associated spin connection is unchanged under this coordinate transformation, since the nonzero components of~\eqref{spinbar2} carry only $r$, $\theta$, and $\phi$ indices, none of which are affected by the transformation $t\to u$. After transforming to EFC, we observe that the radial component $H^{a}{}_{r}$ contains all the terms that diverge at $r=r_{h}$. The coordinate transformation alone is therefore insufficient to remove these divergences; one must also exploit the freedom in the choice of $\psi$ in order to construct a frame that remains finite at the horizon. Noting that
\begin{equation}
\cosh\psi-\varepsilon\sinh\psi=e^{-\varepsilon\psi}, \quad \varepsilon=\pm 1,
\end{equation}
we use this combination to define a regularizing factor that compensates the divergent behavior of the radial components of the coframe. In particular, since the divergent terms scale as $\sqrt{r/(r-r_{h})}$, we require
\begin{equation}
e^{-\varepsilon\psi}=\alpha\sqrt{\frac{r-r_{h}}{r}},
\end{equation}
where $\alpha=\alpha(r)$ is a real-valued function of the radial coordinate that is nowhere vanishing, $\alpha(r)\neq 0$ for all $r\geq r_{h}$, and remains finite throughout the region $r\geq r_{h}$. This leads to
\begin{equation}\label{psih}
\psi=-\varepsilon\ln\left|\alpha\sqrt{\frac{r-r_{h}}{r}}\right|.
\end{equation}
Since the spin connection~\eqref{spinbar2} does not depend explicitly on $\psi$, this choice has no effect on it. Substituting~\eqref{psih} into the coframe~\eqref{fef}, we obtain
\begin{equation}\label{hefF}
\bar{H}^{a}{}_{\mu}=
\begin{pmatrix}
\dfrac{r+\alpha^{2}\left(r-r_{h}\right)}{2r\alpha}
&
-\varepsilon\alpha & 0 & 0
\\[8pt]
\varepsilon\dfrac{r-\alpha^{2}(r-r_{h})}{2r\alpha}
&
\alpha & 0 & 0
\\[8pt]
0 & 0 & r & 0
\\[8pt]
0 & 0 & 0 & r\sin\theta
\end{pmatrix}.
\end{equation}
Every component of the coframe~\eqref{hefF} and of its associated spin connection is manifestly finite at $r=r_{h}$, provided $\alpha(r)\neq 0$, $\alpha(r)$ is finite, and $\chi$ satisfies the regularity condition at the horizon~\eqref{chifinite}. Moreover, when computing the metric through~\eqref{htog}, the standard form of the Schwarzschild metric in EFC is recovered.

The same procedure applied to the proper coframe~\eqref{propert} yields
\begin{equation}\label{propercoor}
\tilde{H}^{a}{}_{\mu}=
\begin{pmatrix}
\dfrac{r+\alpha^{2}(r-r_{h})}{2\alpha r}
&-\varepsilon\alpha & 0 & 0
\\[10pt]
\dfrac{\varepsilon\co_{\theta}\big(\alpha^{2}(r_{h}-r)+r\big)}{2\alpha r}
&
\alpha\co_{\theta} & -r\si_{\theta} & 0
\\[10pt]
-\dfrac{\varepsilon\co_{\phi}\si_{\theta}\big(\alpha^{2}(r_{h}-r)+r\big)}{2\alpha r}
& -\alpha\co_{\phi}\si_{\theta} & -r\co_{\theta}\co_{\phi} & r\si_{\theta}\si_{\phi}
\\[10pt]
-\dfrac{\varepsilon\si_{\theta}\si_{\phi}\big(\alpha^{2}(r_{h}-r)+r\big)}{2\alpha r}
& -\alpha\si_{\theta}\si_{\phi} & -r\co_{\theta}\si_{\phi} & -r\co_{\phi}\si_{\theta}
\end{pmatrix}.
\end{equation}
Here we have used $\chi$ as given in~\eqref{chiSch} for convenience; this choice simplifies the resulting expressions but is not required. All components of this proper frame are finite at $r=r_{h}$. Computing the scalar invariants~\eqref{AVTS} and the torsion scalar~\eqref{TS} for the proper frame~\eqref{propercoor}, and using $\chi=(2n+1)\pi$ so that $\mathscr{A}=0$, the remaining invariants are
\begin{subequations}\label{scalpha}
\begin{equation}
\mathscr{V}=\frac{\left(2(\alpha-1)\alpha+r\alpha'\right)\left(\alpha(r_{h}-2r)+r\left(2+(r_{h}-r)\alpha'\right)\right)}{\alpha^{2}r^{3}},
\end{equation}
\begin{equation}
\mathscr{T}=\frac{\left((\alpha-1)\alpha-r\alpha'\right)\left(\alpha(2r_{h}-r)+r+r(-r_{h}+r)\alpha'\right)}{\alpha^{2}r^{3}},
\end{equation}
\begin{equation}
T=\frac{2(\alpha-1)^{2}\alpha-2\left(\alpha^{2}(r_{h}-r)+r\right)\alpha'}{\alpha^{2}r^{2}}.
\end{equation}
\end{subequations}
These expressions are free of singularities at $r=r_{h}$. Both the proper frame and the torsion scalar invariants are therefore regular at the horizon. TEGR thus admits an analytic coordinate extension\footnote{EFC are horizon-penetrating and allow the Schwarzschild solution to be extended analytically across the event horizon. However, they do not provide the maximal analytic extension of the spacetime, which is obtained in Kruskal--Szekeres coordinates~\cite{wald2010general}.} for its fundamental geometric variables (the frame and spin connection, or equivalently the proper frame), while the corresponding gravitational Lagrangian density remains finite at the horizon.

As an aside, we examine whether the subclass of Schwarzschild TEGR geometries constructed via the horizon-penetrating procedure also admits members satisfying the inertial frame condition~\eqref{inertial}. This cross-check has no bearing on horizon regularity or on the black-hole interpretation, both of which are already established for the present subclass; it merely identifies the subclass for which the proper frame~\eqref{propercoor} is also an inertial comoving frame. Imposing the conditions for $\chi$ in~\eqref{chiSch} and $\psi$ in~\eqref{psi0ASch} fixes the Lorentz sector in such a way that we select a specific timelike congruence. Since we have already chosen $\chi=(2n+1)\pi$, the remaining condition for $\psi$ implies that
\begin{equation}
\cosh^{-1}\left[\frac{\sqrt{r}}{\sqrt{r-r_{h}}}\right]=-\varepsilon\ln\left|\alpha\sqrt{\frac{r-r_{h}}{r}}\right|.
\end{equation}
Solving for $\alpha(r)$ we obtain two branches parametrized by $\varepsilon$,
\begin{subequations}\label{alpha}
\begin{align}
\varepsilon = +1 \; &:\quad
\alpha(r)=\frac{\sqrt{r}}{\sqrt{r}+\sqrt{r_{h}}}, \\[6pt]
\varepsilon = -1 \; &:\quad
\alpha(r)=\frac{\sqrt{r}}{\sqrt{r}-\sqrt{r_{h}}}.
\end{align}
\end{subequations}
The two branches correspond to tetrads adapted to two different radial free-fall congruences: the ingoing case $(\varepsilon=+1)$ and the outgoing case $(\varepsilon=-1)$. In Schwarzschild spacetime (in EFC), these congruences behave differently at the future event horizon located at $r=r_{h}$. Observers following ingoing geodesics cross the horizon, whereas outgoing geodesics cannot be extended across the future event horizon~\cite{wald2010general}. Consequently, only the ingoing branch yields a tetrad that remains regular at $r=r_{h}$, while the outgoing branch necessarily becomes singular.

Returning to Schwarzschild coordinates, keeping $\psi$ as in~\eqref{psih} and $\chi$ as in~\eqref{chiSch} preserves the scalar invariants~\eqref{scalpha}, as follows from the coordinate invariance of the torsion scalar. This confirms that horizon-penetrating coordinates are not essential for obtaining a regular horizon description. Rather, they provide a convenient route to identifying a broader subclass of Schwarzschild TEGR geometries compatible with a black-hole interpretation, all characterized by Lorentz sector functions $\psi$ and $\chi$ that satisfy the horizon regularity requirements.

\subsection{Horizon regularity criterion}

The preceding subsections have established two routes for selecting values of the Lorentz sector functions compatible with a black hole interpretation of the Schwarzschild TEGR geometries: the inertial frame condition and the horizon-penetrating coordinates. We now adopt a third route, in which the regularity of the torsion scalar invariants at $r=r_{h}$ is imposed directly as a mathematical requirement, without invoking any auxiliary physical or geometrical principle. This procedure yields the most general class of Lorentz sector function values compatible with a regular horizon description, with the previous two approaches corresponding to particular members of this class.

We now impose that $\chi$ satisfies~\eqref{chifinite}, and derive the conditions on $\psi$ for the torsion scalar invariants to remain finite at $r=r_{h}$. From~\eqref{torSch}, the necessary conditions for $T$ to remain finite at the horizon are
\begin{subequations}
\begin{equation}\label{con1}
\sqrt{r-r_{h}}\cosh\psi=\beta,
\end{equation}
\begin{equation}\label{con2}
(2r-r_{h})\cosh\psi+2r(r-r_{h})\left[\cosh\psi\right]'=\lambda\sqrt{r-r_{h}},
\end{equation}
\end{subequations}
where $\beta=\beta(r)$ and $\lambda=\lambda(r)$ are real-valued functions of the radial coordinate finite for all $r\geq r_{h}$. From condition~\eqref{con1}, Eq.~\eqref{con2} can be rewritten in terms of $\beta$ as
\begin{equation}
\lambda=\beta+2r\,\beta'.
\end{equation}
This immediately suggests that the derivative of $\beta$ must also remain finite for all $r\geq r_{h}$. Moreover, $\lambda$ is not independent, but is completely determined by $\beta$. The regularity conditions therefore reduce to conditions on the function $\beta$, yielding
\begin{equation}\label{con3}
\beta(r),\beta'(r) \in \mathbb{R}.
\end{equation}
Carrying out a similar analysis for the torsion scalar invariants~\eqref{vSch},~\eqref{tSch}, and~\eqref{aSch}, we find that conditions~\eqref{con1} and~\eqref{chifinite} are the necessary and sufficient conditions for all scalar invariants under consideration to remain finite at $r=r_{h}$. This therefore defines the regular subclass of Schwarzschild TEGR geometries compatible with a regular horizon description, characterized by
\begin{equation}\label{psig}
\psi=\cosh^{-1}\left[\frac{\beta}{\sqrt{r-r_{h}}}\right].
\end{equation}
Since $\cosh(y)\geq 1$, Eq.~\eqref{psig} requires $\beta \geq \sqrt{r-r_{h}}$. The boundary case $\beta=\sqrt{r-r_{h}}$ is however excluded by condition~\eqref{con3}, since $\beta'(r_{h})$ diverges there. Therefore,
\begin{equation}\label{con4}
\beta>\sqrt{r-r_{h}}.
\end{equation}
This strict inequality implies in particular that $\psi= 0$ is excluded from the regular subclass. This is precisely the value most commonly adopted in the TEGR literature when analyzing the Schwarzschild solution  and explains why horizon singularities have repeatedly been encountered.

This result holds for any representative of the regular Schwarzschild TEGR geometries obtained through residual local Lorentz or general coordinate transformations. In particular, the horizon-penetrating values~\eqref{psih} and the inertial comoving values~\eqref{psi0ASch} previously obtained correspond to specific members of the regular subclass characterized by~\eqref{psig}, namely
\begin{subequations}
\begin{align}
\text{horizon-penetrating:} \qquad
\beta &= \frac{\alpha^{2}(r-r_{h})+r}{2\alpha\sqrt{r}},
\\[1ex]
\text{inertial comoving:} \qquad
\beta &= \sqrt{r}.
\end{align}
\end{subequations}
Each admissible choice of the function $\beta$ specifies an inequivalent regular Schwarzschild TEGR geometry. It is important to note that an inertial comoving frame exists for any member of the regular subclass~\eqref{psig}, not only for the special choice $\beta = \sqrt{r}$. For the frame~\eqref{tetrad3} with spin connection~\eqref{spinv3}, specialized to the regular Schwarzschild TEGR subclass with $A_{1} = 1/A_{2}$ as in~\eqref{SchA} and $\psi$ as in~\eqref{psig}, the inertial condition is satisfied by choosing $\psi_{1}$ according to~\eqref{psi0A}. The tetrad then takes the form~\eqref{te3Sch}, while the spin connection~\eqref{spinv3} becomes
\begin{equation} 
\begin{gathered}
\bar{\omega}_{41\theta} = \bar{\omega}_{13\phi}/\sin\theta = \sin\chi \sinh\psi_{2},
\quad
\bar{\omega}_{13\theta} = \bar{\omega}_{14\phi}/\sin\theta = \cos\chi \sinh\psi_{2}, 
\\[1ex]
\bar{\omega}_{42\theta} = \bar{\omega}_{23\phi}/\sin\theta = \sin\chi \cosh\psi_{2}, 
\quad
\bar{\omega}_{23\theta} = \bar{\omega}_{24\phi}/\sin\theta = \cos\chi \cosh\psi_{2}, 
\\[1ex]
\bar{\omega}_{21r} = \psi'_{2},
\quad
\bar{\omega}_{43r} = \chi', \quad
\bar{\omega}_{43\phi} = \cos\theta,
\end{gathered}
\end{equation}
where $\psi_{2} = \psi - \psi_{1}$ is given by
\begin{equation}\label{fpsi}
\psi_{2} = \cosh^{-1}\!\left[\frac{\tilde{\beta}}{\sqrt{r-r_{h}}}\right], \qquad \tilde{\beta}=\frac{\sqrt{r}\beta-\sqrt{r_{h}(\beta^{2}-r+r_{h})}}{\sqrt{r-r_{h}}} .
\end{equation}
This expression parametrizes the spin connection of the general inertial comoving frame in the regular subclass of Schwarzschild TEGR geometries, with the free function $\beta$ subject to~\eqref{con3} and~\eqref{con4}. It also makes explicit an important feature of TG: the inertial structure of the frame and the regularity of the torsion scalar invariants at the horizon are independent conditions. The inertial frame condition constrains the residual Lorentz parameter $\psi_{1}$ through~\eqref{psi0A}, while the horizon regularity condition constrains the Lorentz sector function $\psi$ through~\eqref{psig}. Consequently, every member of the regular subclass admits an inertial comoving frame, and the two requirements can always be satisfied simultaneously for any admissible function $\beta$.

We now specialize the torsion scalar invariants to the Schwarzschild case~\eqref{SchA}, restricted to the regular subclass of geometries characterized by~\eqref{psig}:
\begin{subequations}
\begin{equation}
\mathscr{A}=\frac{4(r_{h}-r)(\chi')^{2}}{9r}
-\frac{16\sin\chi\left(\sin\chi-\beta\sqrt{r}\,\chi'\right)}{9r^{2}},
\end{equation}
\vskip -0.2cm
\begin{dmath}
\mathscr{V}=
\frac{(23r-9r_{h})\beta^{2}+4r^{2}(\beta\beta'+(r_{h}-r){\beta'}^{2})
+32rr_{h}-24r^{2}-9r_{h}^{2}}{4(r_{h}-r+\beta^{2})r^{3}}
+\frac{2\cos(2\chi)}{r^{2}} \\
+\frac{2\cos\chi\left(3\beta+2r\beta'\right)}{r^{5/2}},
\end{dmath}
\vskip -0.2cm
\begin{dmath}
\mathscr{T}=\frac{9r_{h}^{2}+\beta^{2}(9r_{h}-7r)-4\beta r^{2}\beta'
+4r\left(-4r_{h}+2r+r(r-r_{h}){\beta'}^{2}\right)}{4r^{3}(r-r_{h}-\beta^{2})}
+\frac{\cos\chi\left(3\beta-2r\beta'\right)}{r^{5/2}}
+\frac{2\beta\sin\chi\,\chi'}{r^{3/2}}
+\frac{(r-r_{h}){\chi'}^{2}}{r},
\end{dmath}
\vskip 0.2cm
\begin{equation}
T=-\frac{4}{r^{2}}+\frac{4\beta r\sin\chi\,\chi'
-2\cos\chi\left(\beta+2r\beta'\right)}{r^{5/2}}.
\end{equation}
\end{subequations}
All displayed torsion scalar invariants are manifestly regular at $r=r_{h}$, and the same conclusion extends to higher-order torsion scalar invariants. These observations clarify why a black hole interpretation has often been regarded as unattainable in the TEGR literature: the commonly adopted value $\psi=0$ lies outside the regular subclass. The Schwarzschild TEGR geometries associated with this value of the Lorentz sector function exhibit genuine divergences of the torsion scalar invariants at $r=r_{h}$ and therefore do not admit a black hole interpretation~\cite{obukhov2003metric,coley2025black,lopez2025black}.

The three complementary approaches developed in this section converge on a single conclusion: the Schwarzschild solution in TEGR admits a consistent black-hole interpretation precisely for the class of Lorentz sector function values characterized by~\eqref{chifinite} and~\eqref{psig}. The inertial frame condition and the horizon-penetrating construction identify particular members of this class, while the horizon regularity criterion provides its most general characterization.

\section{Discussion}

The Schwarzschild solution in TEGR carries a structural feature that is invisible at the metric level: the values of the Lorentz sector functions determine whether the teleparallel geometry is well-defined at the horizon, since the same metric admits both regular and singular teleparallel realizations. This feature has no counterpart in sRG, where the metric tensor specifies the geometry completely and curvature scalar invariants are uniquely determined by it.

The horizon divergences of torsion scalar invariants in the Schwarzschild solution of TEGR have been a persistent feature of the literature~\cite{obukhov2003metric,golovnev2024static,coley2025black,lopez2025black}, and have often been interpreted as an obstruction intrinsic to the teleparallel description. The present analysis offers a different interpretation. Within the family of Schwarzschild TEGR geometries, the value $\psi = 0$ has been the standard choice, both because it is the simplest choice and because it trivially satisfies the requirement that all torsion components vanish as $r_{h}\to 0$. This value corresponds to $\beta = \sqrt{r - r_h}$, which lies outside the regular subclass~\eqref{psig}; the divergences encountered in the literature are therefore genuine features of the particular subclass under study, not of the Schwarzschild solution in TEGR as a whole. The distinction is conceptual rather than computational: in the absence of a structural distinction between the Lorentz sector and the residual Lorentz freedom, all members of the family appear physically indistinguishable at the metric level, and the regular subclass remains hidden among the formally legitimate choices. The existence of this regular subclass was established through three independent approaches: (i) the inertial frame condition, (ii) horizon-penetrating coordinates, and (iii) the horizon regularity criterion. Although conceptually distinct, all three approaches converge on the same family of Lorentz sector functions, providing mutually consistent and independent confirmation of the regularity conditions~\eqref{chifinite} and~\eqref{psig}.

The framework that makes this partition transparent is the distinction between the metric and the Lorentz sector as separate carriers of geometric information. In sRG, the metric tensor $g_{\mu\nu}$ specifies the geometry completely: curvature scalar invariants are uniquely determined by the metric, and a divergence of such an invariant signals a feature of the spacetime itself. In TG, by contrast, the geometry is specified by the pair $(h^a{}_\mu, \omega^a{}_{b\mu})$, and the Lorentz sector encodes geometric information that is absent at the metric level. Torsion scalar invariants therefore probe a richer structure than curvature invariants do: they depend not only on the metric but also on the Lorentz sector, and their values can differ across inequivalent teleparallel geometries that share the same metric. A divergence of a torsion scalar invariant is therefore a feature of a particular teleparallel geometry rather than of the underlying metric solution. The partition of the Schwarzschild family is a direct expression of this fact: the regularity or singularity of the torsion scalar invariants at the horizon is determined entirely by the Lorentz sector, independently of the metric.

A related conceptual issue concerns the role of the proper frame in TG. The proper frame, defined by the vanishing of the spin connection $\omega^{a}{}_{b\mu} = 0$, is widely taken to be a frame in which inertial contributions are absent~\cite{krvsvsak2015spin,krvsvsak2019teleparallel,aldrovandi2008inertia,
lucas2009regularizing}. We have shown that this identification does not hold: the local Lorentz transformation that brings the spin connection to zero transfers the inertial content of the geometry to the tetrad, so that the equation of motion in the proper frame still contains the contortion tensor and does not reduce to its special-relativistic free-fall form. The physically relevant criterion for the absence of inertial effects is instead the inertial condition $\Sigma_{ab} = 0$, which aligns the tetrad transport with the equation of motion along the observer's worldline. The proper frame becomes free of inertial effects only when this condition is imposed in addition to $\omega^{a}{}_{b\mu} = 0$, defining the more restrictive inertial proper frame. This misidentification has a direct bearing on the interpretation of the Lorentz sector dependence of the torsion scalar.

A further consequence of this structural distinction concerns the interpretation of the Lorentz sector dependence of the torsion scalar. It has been claimed in the literature that the torsion scalar carries spurious inertial contributions, and that choosing the appropriate spin connection serves to remove them~\cite{krvsvsak2015spin,krvsvsak2019teleparallel}. The torsion scalar is however a Lorentz-invariant quantity and is therefore completely insensitive to any local Lorentz transformation, including those that implement the inertial condition~\eqref{inertial}. Computing the torsion scalar in an inertial comoving frame leaves it unchanged, which means it carries no inertial contributions in any sense: it cannot be contaminated by inertia, and no choice of spin connection can purify it of inertial effects, because there are none to remove. What the spin connection actually controls is the Lorentz sector of the teleparallel geometry. Different values of $\psi$ and $\chi$ select inequivalent teleparallel geometries, and this is why the torsion scalar depends on them---not because they encode inertia, but because they encode geometry.

\section{Conclusions}

The main results of this work are twofold. First, starting from the general class of Schwarzschild TEGR geometries, we derived the necessary and sufficient conditions~\eqref{chifinite} and~\eqref{psig} for the regularity of all torsion scalar invariants at the horizon. These conditions define a distinguished subclass of Schwarzschild TEGR geometries admitting a consistent black-hole interpretation. Consequently, the Schwarzschild solution in TEGR corresponds to a family of inequivalent teleparallel geometries parametrized by the Lorentz sector functions $\psi$ and $\chi$, which partitions into a regular subclass and a complementary subclass in which the torsion scalar invariants diverge at the horizon and no such interpretation is available. Second, we have shown that every member of the regular subclass admits an inertial comoving frame, explicitly parametrized by~\eqref{fpsi} for any admissible function $\beta$ subject to~\eqref{con3} and~\eqref{con4}. This establishes that horizon regularity and the inertial structure of the frame are independent and simultaneously satisfiable conditions in TG.

The specific regularity conditions~\eqref{chifinite} and \eqref{psig} are tied to the structure of the Schwarzschild solution in TEGR, where the Lorentz sector is unaffected by the FE. In other teleparallel theories, such as $F(T)$ gravity and NGR, the FE generally depend on the Lorentz sector and constrain it dynamically, and the regular subclass identified here may not extend directly to those settings. The conceptual framework, however, does extend: horizon singularities of torsion scalar invariants in any teleparallel theory should be diagnosed only after the Lorentz sector freedom available in the theory has been identified and controlled. Singular behaviour on a horizon should not, by itself, be taken as evidence against a black hole interpretation without first establishing what subclass of teleparallel geometries the singularity belongs to.

The present work invites several extensions. The most immediate is the analysis of axially symmetric TEGR geometries, where the Kerr solution plays a role analogous to that of the Schwarzschild solution. It is natural to ask whether the freedom in the Lorentz sector similarly partitions Kerr TEGR geometries into subclasses distinguished by the regularity properties of their torsion scalar invariants at the horizon. A second extension concerns teleparallel theories in which the FE constrain the Lorentz sector partially: in such cases, the constrained Lorentz sector may or may not intersect the regular subclass, and the diagnosis of horizon behaviour turns on whether the dynamics is compatible with regularity. Both extensions sharpen the diagnostic role of the Lorentz sector in teleparallel theories, and clarify under what conditions horizon singularities reflect genuine geometric pathology rather than the structure of a particular subclass of geometries.

\section*{Acknowledgement}

D. F. L. acknowledges support from the Department of Mathematics and Statistics at Dalhousie University, Canada. A. A. C. and R. J. vdH. are supported by the Natural Sciences and Engineering Research Council of Canada (NSERC).  

\printbibliography

@article{hayashi1979new,
  author       = {Hayashi, K. and Shirafuji, T.},
  title        = {New General Relativity},
  journaltitle = {Physical Review D},
  shortjournal = {Phys. Rev. D},
  volume       = {19},
  number       = {12},
  pages        = {3524--3553},
  year         = {1979},
  doi          = {10.1103/PhysRevD.19.3524}
}

@article{obukhov2003metric,
  author       = {Obukhov, Y. N. and Pereira, J. G.},
  title        = {Metric-Affine Approach to Teleparallel Gravity},
  journaltitle = {Physical Review D},
  shortjournal = {Phys. Rev. D},
  volume       = {67},
  number       = {4},
  pages        = {044016},
  year         = {2003},
  doi          = {10.1103/PhysRevD.67.044016},
  eprint       = {gr-qc/0212080},
  eprinttype   = {arXiv}
}

@article{golovnev2024static,
  author       = {Golovnev, A. and Semenova, A. N. and Vandeev, V. P.},
  title        = {Static Spherically Symmetric Solutions in New General Relativity},
  journaltitle = {Classical and Quantum Gravity},
  shortjournal = {Class. Quantum Grav.},
  volume       = {41},
  number       = {5},
  pages        = {055009},
  year         = {2024},
  doi          = {10.1088/1361-6382/ad1f1c},
  eprint       = {2305.03420},
  eprinttype   = {arXiv},
  eprintclass  = {gr-qc}
}

@article{bahamonde2023teleparallel,
  author       = {Bahamonde, S. and Dialektopoulos, K. F. and Escamilla-Rivera, C. and Farrugia, G. and Gakis, V. and Hendry, M. and Hohmann, M. and Said, J. L. and Mifsud, J. and Di Valentino, E.},
  title        = {Teleparallel Gravity: From Theory to Cosmology},
  journaltitle = {Reports on Progress in Physics},
  shortjournal = {Rep. Prog. Phys.},
  volume       = {86},
  number       = {2},
  pages        = {026901},
  year         = {2023},
  doi          = {10.1088/1361-6633/ac9cef},
  eprint       = {2106.13793},
  eprinttype   = {arXiv},
  eprintclass  = {gr-qc}
}

@article{mcnutt2023frame,
  author       = {McNutt, D. D. and Coley, A. A. and van den Hoogen, R. J.},
  title        = {A Frame-Based Approach to Computing Symmetries with Nontrivial Isotropy Groups},
  journaltitle = {Journal of Mathematical Physics},
  shortjournal = {J. Math. Phys.},
  volume       = {64},
  number       = {3},
  pages        = {032502},
  year         = {2023},
  doi          = {10.1063/5.0134815},
  eprint       = {2302.11493},
  eprinttype   = {arXiv},
  eprintclass  = {gr-qc}
}

@book{aldrovandi2012teleparallel,
  author    = {Aldrovandi, R. and Pereira, J. G.},
  title     = {Teleparallel Gravity: An Introduction},
  year      = {2012},
  publisher = {Springer},
  series    = {Fundamental Theories of Physics},
  volume    = {173},
  doi       = {10.1007/978-94-007-5143-9},
  isbn      = {978-94-007-5142-2}
}

@article{Emtsova:2021ehh,
  author       = {Emtsova, E. D. and Kr{\v{s}}{\v{s}}{\'a}k, M. and Petrov, A. N. and Toporensky, A. V.},
  title        = {On Conserved Quantities for the Schwarzschild Black Hole in Teleparallel Gravity},
  journaltitle = {The European Physical Journal C},
  shortjournal = {Eur. Phys. J. C},
  volume       = {81},
  number       = {8},
  pages        = {743},
  year         = {2021},
  doi          = {10.1140/epjc/s10052-021-09505-x},
  eprint       = {2105.13312},
  eprinttype   = {arXiv},
  eprintclass  = {gr-qc}
}

@book{misner1973gravitation,
     author    = {Misner, C. W. and Thorne, K. S. and Wheeler, J. A.},
     title     = {Gravitation},
     publisher = {W. H. Freeman},
     address   = {San Francisco},
     year      = {1973},
     isbn      = {978-0-7167-0344-0}
   }

@book{hawking2023large,
  author    = {Hawking, S. W. and Ellis, G. F. R.},
  title     = {The Large Scale Structure of Space-Time},
  year      = {1973},
  publisher = {Cambridge University Press},
  isbn      = {978-0-521-09906-6}
}

@incollection{krvsvsak2024teleparallel,
  author       = {Krššák, M.},
  title        = {Teleparallel Gravity, Covariance and Their Geometrical Meaning},
  booktitle    = {In Tribute to Ruben Aldrovandi},
  year         = {2024},
  publisher    = {Editora Livraria da Física},
  location     = {São Paulo},
  eprint       = {2401.08106},
  eprinttype   = {arXiv},
  eprintclass  = {gr-qc}
}

@article{coley2024spherically,
  author       = {Coley, A. A. and Landry, A. and van den Hoogen, R. J. and McNutt, D. D.},
  title        = {Spherically Symmetric Teleparallel Geometries},
  journaltitle = {The European Physical Journal C},
  shortjournal = {Eur. Phys. J. C},
  volume       = {84},
  number       = {3},
  pages        = {334},
  year         = {2024},
  doi          = {10.1140/epjc/s10052-024-12615-0},
  eprint       = {2402.07238},
  eprinttype   = {arXiv},
  eprintclass  = {gr-qc}
}

@article{coley2025black,
  author       = {Coley, A. A. and Layden, N. T. and López, D. F.},
  title        = {On Black Holes in Teleparallel Torsion Theories of Gravity},
  journaltitle = {General Relativity and Gravitation},
  shortjournal = {Gen. Relativ. Gravit.},
  volume       = {57},
  number       = {3},
  pages        = {59},
  year         = {2025},
  doi          = {10.1007/s10714-025-03359-3},
  eprint       = {2503.13215},
  eprinttype   = {arXiv},
  eprintclass  = {gr-qc}
}

@article{lopez2025black,
  author       = {L{\'o}pez, D. F. and Coley, A. A. and van den Hoogen, R. J.},
  title        = {On Black Holes in New General Relativity},
  journaltitle = {Classical and Quantum Gravity},
  shortjournal = {Class. Quantum Grav.},
  volume       = {42},
  number       = {24},
  pages        = {245011},
  year         = {2025},
  doi          = {10.1088/1361-6382/ad5f34},
  eprint       = {2508.20314},
  eprinttype   = {arXiv},
  eprintclass  = {gr-qc}
}

@article{weatherall2024general,
  author       = {Weatherall, J. O. and Meskhidze, H.},
  title        = {Are General Relativity and Teleparallel Gravity Theoretically Equivalent?},
  journaltitle = {Philosophy of Physics},
  shortjournal = {Phil. Phys.},
  volume       = {3},
  number       = {1},
  pages        = {6},
  year         = {2025},
  doi          = {10.31389/pop.152},
  eprint       = {2406.15932},
  eprinttype   = {arXiv},
  eprintclass  = {physics.hist-ph}
}

@article{van2024teleparallel,
  author       = {van den Hoogen, R. J. and Forance, H.},
  title        = {Teleparallel Geometry with Spherical Symmetry: The Diagonal and Proper Frames},
  journaltitle = {Journal of Cosmology and Astroparticle Physics},
  shortjournal = {J. Cosmol. Astropart. Phys.},
  volume       = {2024},
  number       = {11},
  pages        = {033},
  year         = {2024},
  doi          = {10.1088/1475-7516/2024/11/033},
  eprint       = {2408.13342},
  eprinttype   = {arXiv},
  eprintclass  = {gr-qc}
}

@article{krvsvsak2019teleparallel,
  author       = {Kr{\v{s}}{\v{s}}{\'a}k, M. and van den Hoogen, R. J. and Pereira, J. G. and B{\"o}hmer, C. G. and Coley, A. A.},
  title        = {Teleparallel Theories of Gravity: Illuminating a Fully Invariant Approach},
  journaltitle = {Classical and Quantum Gravity},
  shortjournal = {Class. Quantum Grav.},
  volume       = {36},
  number       = {18},
  pages        = {183001},
  year         = {2019},
  doi          = {10.1088/1361-6382/ab2eaa},
  eprint       = {1810.12932},
  eprinttype   = {arXiv},
  eprintclass  = {gr-qc}
}

@article{krvsvsak2015spin,
  author       = {Kr{\v{s}}{\v{s}}{\'a}k, M. and Pereira, J. G.},
  title        = {Spin Connection and Renormalization of Teleparallel Action},
  journaltitle = {The European Physical Journal C},
  shortjournal = {Eur. Phys. J. C},
  volume       = {75},
  number       = {11},
  pages        = {519},
  year         = {2015},
  doi          = {10.1140/epjc/s10052-015-3803-7},
  eprint       = {1504.07683},
  eprinttype   = {arXiv},
  eprintclass  = {gr-qc}
}

@book{kobayashi1996foundations,
  author    = {Kobayashi, S. and Nomizu, K.},
  title     = {Foundations of Differential Geometry, Volume 2},
  year      = {1996},
  publisher = {John Wiley \& Sons},
  volume    = {2},
  isbn      = {978-0471157335}
}

@book{wald2010general,
  author    = {Wald, R. M.},
  title     = {General Relativity},
  year      = {2010},
  publisher = {University of Chicago Press},
  isbn      = {978-0-226-87033-5}
}

@article{krvsvsak2016covariant,
  author       = {Kr{\v{s}}{\v{s}}{\'a}k, M. and Saridakis, E. N.},
  title        = {The Covariant Formulation of {$F(T)$} Gravity},
  journaltitle = {Classical and Quantum Gravity},
  shortjournal = {Class. Quantum Grav.},
  volume       = {33},
  number       = {11},
  pages        = {115009},
  year         = {2016},
  doi          = {10.1088/0264-9381/33/11/115009},
  eprint       = {1510.08432},
  eprinttype   = {arXiv},
  eprintclass  = {gr-qc}
}

@article{maluf2013teleparallel,
  author       = {Maluf, J. W.},
  title        = {The Teleparallel Equivalent of General Relativity},
  journaltitle = {Annalen der Physik},
  shortjournal = {Ann. Phys. (Berlin)},
  volume       = {525},
  number       = {5},
  pages        = {339--357},
  year         = {2013},
  doi          = {10.1002/andp.201200272},
  eprint       = {1303.3897},
  eprinttype   = {arXiv},
  eprintclass  = {gr-qc}
}

@article{aldrovandi2008inertia,
  author  = {Aldrovandi, R. and Lucas, T. G. and Pereira, J. G.},
  title   = {Inertia and Gravitation in Teleparallel Gravity},
  journal = {General Relativity and Gravitation},
  volume  = {41},
  pages   = {2173--2185},
  year    = {2009},
  doi     = {10.1007/s10714-008-0730-6},
note = {arXiv:0812.0034 [gr-qc]}
}

@article{lucas2009regularizing,
  author       = {Lucas, T. G. and Obukhov, Y. N. and Pereira, J. G.},
  title        = {Regularizing Role of Teleparallelism},
  journaltitle = {Physical Review D},
  shortjournal = {Phys. Rev. D},
  volume       = {80},
  number       = {6},
  pages        = {064043},
  year         = {2009},
  doi          = {10.1103/PhysRevD.80.064043},
  eprint       = {0909.2418},
  eprinttype   = {arXiv},
  eprintclass  = {gr-qc}
}

@article{maluf2007reference,
  author       = {Maluf, J. W. and Faria, F. F. and Ulhoa, S. C.},
  title        = {On Reference Frames in Spacetime and Gravitational Energy in Freely Falling Frames},
  journaltitle = {Classical and Quantum Gravity},
  shortjournal = {Class. Quantum Grav.},
  volume       = {24},
  number       = {10},
  pages        = {2743--2754},
  year         = {2007},
  doi          = {10.1088/0264-9381/24/10/017},
  eprint       = {0704.0986},
  eprinttype   = {arXiv},
  eprintclass  = {gr-qc}
}

\end{document}